\documentclass[12pt,preprint2]{aastex}

\slugcomment{}

\shorttitle{Cloud fragmentation and proplyd-like features}
\shortauthors{De Marco et al.}

\begin{document}


\title{Cloud fragmentation and proplyd-like features \\
in \ion{H}{2} regions imaged by HST$^1$}


\author{Orsola De Marco\altaffilmark{2}, C.R. O'Dell\altaffilmark{3},
Pamela Gelfond\altaffilmark{4}, R. H. Rubin\altaffilmark{5,6}, \&
S.C.O. Glover\altaffilmark{2}
}

\altaffiltext{1}
{Based on observations made with the NASA/ESA {\it Hubble Space Telescope}
(HST) obtained at the Space Telescope Science Institute, which is operated
by the Association of Universities for Research in Astronomy (AURA), 
Inc., under NASA contract NAS5-26555.}
\altaffiltext{2}{Astrophysics Department, American Museum of Natural History, \\
Central Park West at 79$^{th}$ Street, New York, NY 10024 \\
orsola@amnh.org}
\altaffiltext{3}{Department of Physics and Astronomy, Vanderbilt University\\ 
Box 1807-B, Nashville, TN 37235\\ 
cr.odell@vanderbilt.edu}
\altaffiltext{4}{Department of Physics, Yale University\\
217 Prospect Street, New Haven, CT 06511\\
pamela.gelfond@yale.edu}
\altaffiltext{5}{NASA Ames Research Center\\
Moffett Field, CA 94035-1000\\
rubin@cygnus.arc.nasa.gov}
\altaffiltext{6}{Orion Enterprises, M.S. 245-6, Moffett Field, CA 94035-1000}


%


\begin{abstract}
We have analyzed HST ACS and WFPC2 new and archival images of eight \ion{H}{2} regions
to look for new proto-planetary disks (proplyds) similar
to those found in the Orion Nebula. We find a wealth of features similar in size
(though many are larger) to
the bright cusps around the Orion Nebula proplyds. None of them, however, contains a
definitive 
central star. From this, we deduce that the new cusps may not be proplyds, but
instead are fragments of molecular cloud material.
Out of {\it all} the features found in the eight \ion{H}{2} regions examined,
only one, an apparent edge-on silhouette in M~17,
may have a central star. This feature might join the small number of 
{\it bona fide}
proplyds
found outside the Orion Nebula, in M~8, M~20 and possibly in M~16. 
In line with the results found recently by 
Smith et al. (2005), 
the paucity of proplyds outside the Orion Nebula, 
may 
be explained by
their transient nature as well as by the specific environmental conditions under which
they can be observed.

Several fragments are seen as dark silhouettes against a bright background. We
have re-analyzed those found in IC~2944 by \citet{Rei+03} and found new, similar
ones in M~16. None of these fragments contains a central
star and we exclude that they are disks. \citet{Rei+03} 
concluded that the IC~2944 silhouettes are not
star-forming. We argue here that their assumption of a constant optical
depth for these fragments is not physical and that it is more likely
that these fragments are star forming, a condition that is supported, although
not proven, by their
shapes and distributions. 
The process of cloud fragmentation {\it and}
photo-evaporation produces a large number of small fragments, 
while the size hierarchy expected in a photo-evaporative environment would
not favor small fragments. The size distributions observed
will constrain any future theories of cloud fragmentation.

One bright micro-jet candidate is found in M~17 protruding from a large, limb-brightened
fragment. A second, larger jet-like feature, similar in shape and size to a Herbig-Haro jet
is found in Pismis~24.
No central star appears to be associated with either of these jet candidates.

\end{abstract}



\section{Introduction}

The basic outlines of the processes of star formation are widely understood.
We see the raw material for star formation in molecular clouds, and the
end-product is seen in pre-main sequence and young massive stars on the main
sequence.
The hierarchical fragmentation of the molecular clouds and the formation of
nascent
stars has been elaborated from both ends of the process. At the beginning of
the
process one sees density concentrations within the molecular clouds, with
sizes ranging down to the resolution of the methods for
measuring
them \citep{Van+01}.
At the end of the process one sees young stellar objects (YSOs) such
as
the T Tauri stars and stars shrouded with varying amounts of obscuring
opacity.
A special class of YSO is the proplyds (a word designed to replace the longer ``PROto-PLanetarY Disks";
\citealp{ODe+93}), 
which are YSO's found in or near an
\ion{H}{2}
region, where the conditions for their observation are different from
regions
without a nearby ionizing star.  The proplyds are easily visible because
some are
directly illuminated by the star(s) producing the \ion{H}{2} region and hence have
bright
photo-ionized cusps facing the source of ionizing radiation. Other proplyds
are not
photo-ionized, but their dust component allows them to be seen in silhouette
against
the background emission of the \ion{H}{2} region, while others demonstrate both
properties.  This means that it is possible
to detail the nature of this class of YSO by
direct
imaging.

The early part of cloud fragmentation involves formation of large clumps which
include the classical ``Bok globules" \citep{BR47}, some of which contain YSO's and 
others which may do
so in the future.  The physical size of such objects is about 0.35 parsecs \citep{CYH91}.
The smallest 
fragmentation products will look like proplyds in that they will have outer portions
smoothly sculpted by photo-evaporation of ionized material and cores that appear
in silhouette.  It is difficult to discern between a true proplyd, with an associated
young star, and a compact fragment of similar appearance. The best studied
region in the search for proplyds is the Orion Nebula with its embedded star cluster. The
Orion Nebula Cluster is the closest cluster of stars
containing an \ion{H}{2} region \citep{ODe01}. There we see circumstellar material with diameters
as large as 1730 AU (0.08 parsecs) and as small as the resolution of the Hubble
Space Telescope's (HST) cameras (about 50 AU).

The search for proplyds in other \ion{H}{2} regions is more difficult because of
their
greater distances. This makes it difficult to discriminate between a true proplyd,
which by its classification as a YSO includes a nascent star, and fragmented
portions of a molecular cloud that are subject to the same processes of
illumination.

The search for proplyds has driven several studies 
\citep{Bra+00,Sta+97,Ste+98,Bal+98,Smi+03}, but results have been slow-coming.
We know that in the case of the true proplyds, stars have formed and the
circumstellar material is the remainder of the star's placental material.
Cloud fragments of larger size
found in \ion{H}{2} regions may or may not survive the hostile 
photo-evaporation
environment long enough to produce a star. 
Several studies reported the detection of cloud fragments, some of which were tentatively interpreted
as proplyd structures (e.g., \citealp{Smi+03}), while others were thought {\it not} to be forming 
stars (e.g., \citealp{Rei+03}). Never has a {\it bona fide} family of proplyds similar to those found in the
Orion Nebula been found in another \ion{H}{2} region.

The recent work of 
\citet{Smi+05b} may 
provide the answer to the proplyds' elusive nature.
They found that the Orion Nebula proplyds are concentrated 
in the middle of the region, near the {\it very youngest} cluster stars, with ages of less than 0.5~Myr. 
Farther from the center, the number of proplyds
diminishes, while the number of naked stars with
low mass disks (detected via mid-IR emission) increases. They concluded that the typical proplyds
(star, disk and cusp) are very short-lived, so that after about 0.5~Myr only their descendants (stars with light disks)
remain. 
If this interpretation is correct, it might
explain the rarity of proplyds in other \ion{H}{2} regions: to see proplyds we need nearby
\ion{H}{2} regions with very bright, extremely young stars.

In this investigation we increase the number of proplyd-like (and larger) 
fragments observed in \ion{H}{2} regions.
By determining the size hierarchy and morphological characteristics
of these clumps of gas and dust in the context of those found in the literature,
we provide new data to constrain theories of
cloud fragmentation that may or may not produce stars. 

In Section~\ref{sec:observations}, we discuss the 
observations.
In 
Section~\ref{sec:thestarclusters}, we review the sample clusters' properties from 
the literature;
we survey 
the large scale Digital Sky Survey (DSS),
and we classify interesting features within
the Advanced Camera for Surveys (ACS) and Wide Field and Planetary Camera 2 (WFPC2) 
fields of view (FOVs).
In Section~\ref{sec:analysis}, we compare the regions and interpret their features
within the context of the Orion Nebula proplyds and proplyd-like features encountered elsewhere.
We summarize in Section~\ref{sec:summary}.

\section{Observations}
\label{sec:observations}

The observations presented in this paper were taken by HST ACS/WFI (FOV=202$\times$202~arcsec$^2$) 
and WFPC2 (FOV=150$\times$150~arcsec$^2$) Cycle 13 program
GO9857 (PI: O. De Marco). In addition we have analyzed archival images from 
the ACS Early Release Observations of M~17
(program GO8992; PI: H. C. Ford) and the
WFPC2 images of IC~2944 and M~16 acquired by programs GO7381 (PI: B. Reipurth) and GO9091 (PI: J. J. Hester),
respectively.
A list of the observations can be found in Table~\ref{tab:observations}.
The RA and Dec in Column 2 are the coordinates of the chip centers, measured in 
IRAF\footnote{{\sc iraf} is distributed by the National Optical Astronomy Observatories,
which are operated by the Association of Universities for Research
in Astronomy, Inc., under cooperative agreement with the National
Science Foundation.}
\citep{Tod86,Tod93} using the
{\sc ds9} \citep{JM03}
display, rather than from the image header.
The position angle in Column 3 is the angle between the north direction and
the y-axis of the ACS/WFC or WFPC2 chip assemblies. The ACS angle is obtained by adding 180~deg
to the V3 angle from the image header, while the WFPC2 angle is obtained by adding 224.86~deg.
All images were calibrated by the on-the-fly calibration pipeline. This was
deemed sufficient for the 
purpose
of the current investigation. 

DSS images (POSS2/UKSTU, taken through the red filter, 
with a plate-scale of 1.01 arcsec/pix; \citealp{Rei+91}) 
of the \ion{H}{2} regions, with an overlay of the HST apertures, are displayed in
Figs.~\ref{fig:ngc6530_apertures} to \ref{fig:m17_apertures}, while cluster and region parameters are listed
in Table~\ref{tab:clusters}. The cluster coordinates are from the SIMBAD database
\footnote{The SIMBAD database,
operated at CDS, Strasbourg, France.}.
On these images we also mark the O stars.
The WFPC2 apertures for the 6 visits of
program GO9857 are also marked on the DSS images, although they are not presented in this
paper due to the lack of interesting interstellar features.

\section{The ionizing stars and cloud fragments within the \ion{H}{2} regions}
\label{sec:thestarclusters}

In this Section we review the observed \ion{H}{2} regions' 
properties (Table~\ref{tab:clusters}) as well as their stellar
content (Table~\ref{tab:stars}). From the large-scale DSS images we determine
the apparent size and general appearance of the \ion{H}{2} regions 
as well as the distribution of O stars 
in relation to the position of the ACS or WFPC2 apertures
(Figs.~\ref{fig:ngc6530_apertures} to \ref{fig:m17_apertures}).

Feature numbers mentioned in the subsections below are those
marked on the respective images (Figs.~\ref{fig:visit2} to \ref{fig:m17})
and refer to the IDs in Table~\ref{tab:features}. One or two dimensions in AU are
listed for every feature, where the distances in Table~\ref{tab:clusters} have been used
to convert angular into linear scales. The uncertainty on the measurements is $\sim$5\%. 
Features discussed explicitly, but too small
to be seen in Figs.~\ref{fig:visit2} to \ref{fig:m17} are also 
shown as thumbnails in Fig.~\ref{fig:proplydstamps}\footnote{While we
do not attempt a precise morphological classification in this figure, 
the features are approximately sorted by type:
attached cusps are followed by semi-attached ones, then detached ones. These
are followed by tiny silhouettes called nodules larger circular silhouettes and
an assortment of more complex fragments.}. 
All discrete features in the HST images smaller than $\sim$10\,000~AU
(most of which bear a resemblance to the Orion Nebula proplyds),
are classified as either {\it cusps} or {\it silhouettes}.
Cusps are limb-brightened, generally on one side. Silhouettes
are not limb-brightened. Cusps and silhouettes can be attached to a large fragment 
or a ridge. When this is so they look like fingers or protuberances.
We also list fragments larger than 10\,000~AU in at least one direction
(and call them {\it fragments} in Table~\ref{tab:features},
although we also use the word in the text
for features in general). These tend to be large and complex in shape
with edges can be sharp, fuzzy, limb-brightened or a mix of the above.

\subsection{NGC~6530}
\label{ssec:ngc6530}

\begin{figure}
\plotone{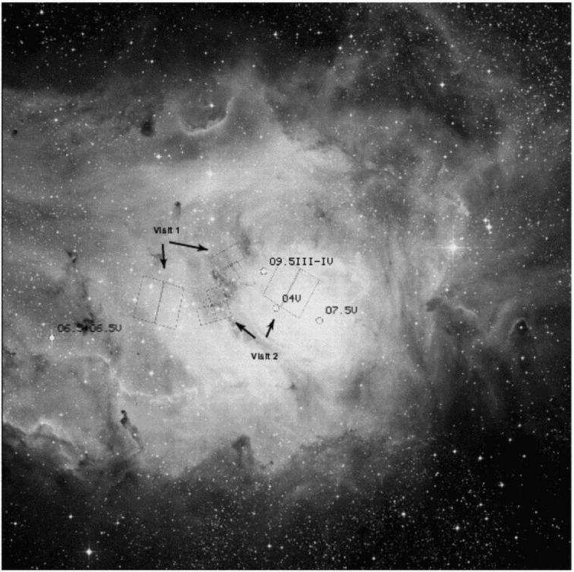}
\caption{A 45x45~arcmin$^2$ DSS/POSSII (red filter) image of 
NGC~6530 (Visits 1 and 2), with an overlay of the ACS and 
WFPC2 apertures used. North is toward the top, east toward the left. 
The O stars are marked (see Table~\ref{tab:stars}). 
\label{fig:ngc6530_apertures}}
\end{figure}

\begin{figure}
\plotone{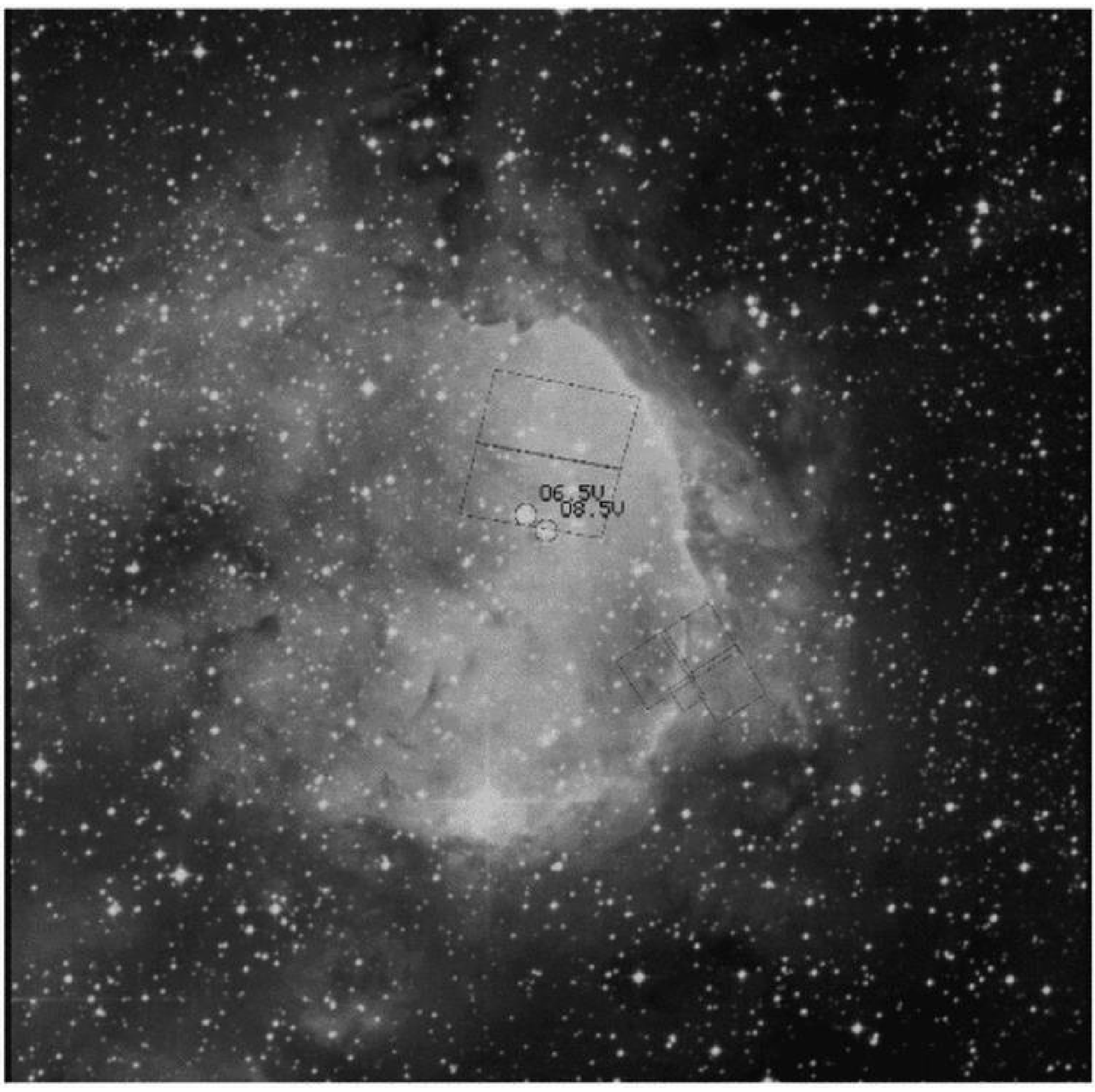}
\caption{A 25x25~arcmin$^2$ DSS/POSSII (red filter) 
image of NGC~3324, with an overlay of the ACS and
WFPC2 apertures used. North is toward the top, east toward the left. 
The O stars are marked (see Table~\ref{tab:stars}).
\label{fig:ngc3324_apertures}}
\end{figure}

\begin{figure}
\plotone{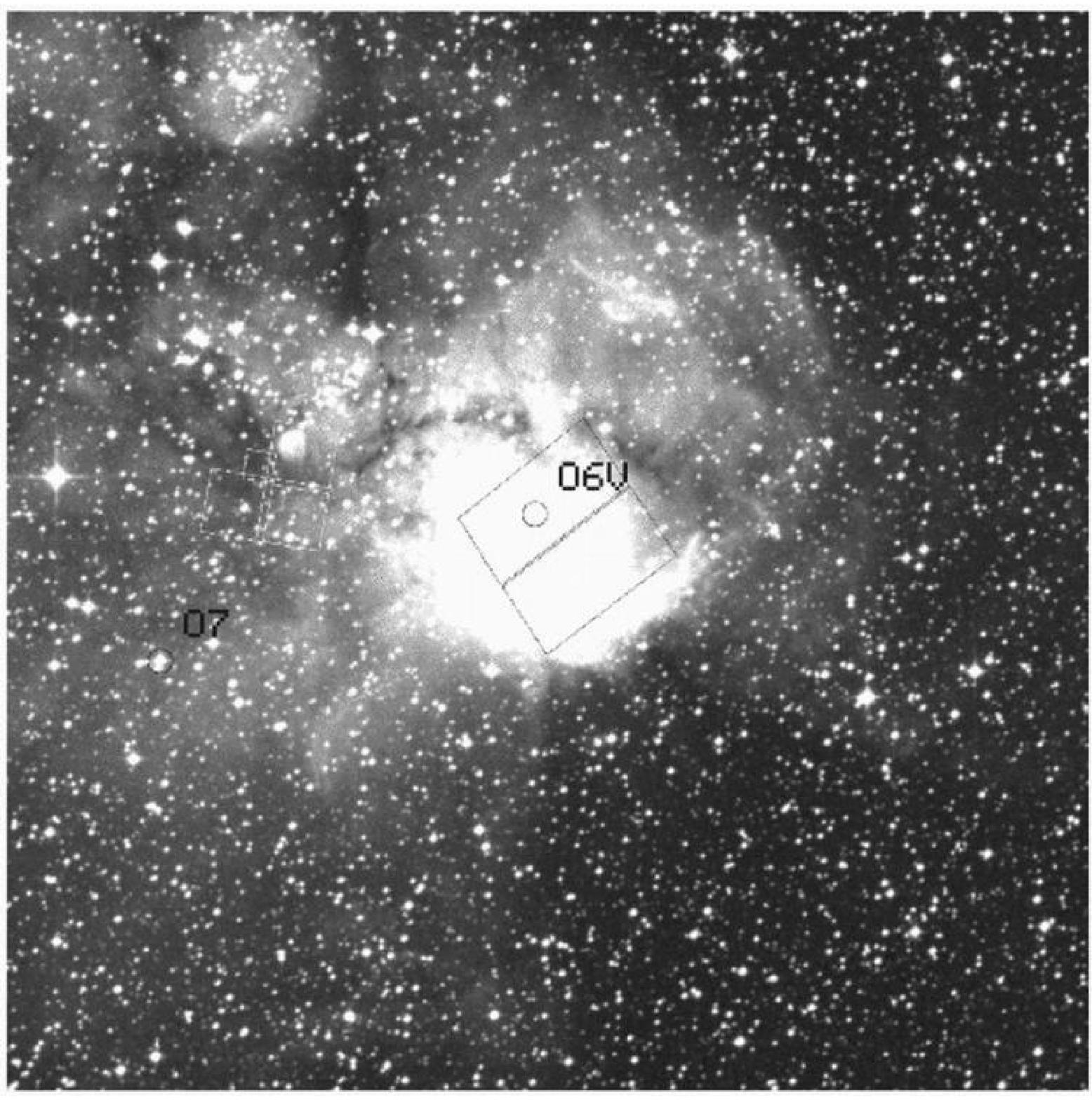}
\caption{A 25x25~arcmin$^2$
DSS/POSSII (red filter) image of NGC~2467 with an overlay of the ACS and WFPC2
apertures used. North is toward the top, east toward the left. 
The O stars are marked (see Table~\ref{tab:stars}).
\label{fig:ngc2467_apertures}}
\end{figure}

\begin{figure}
\plotone{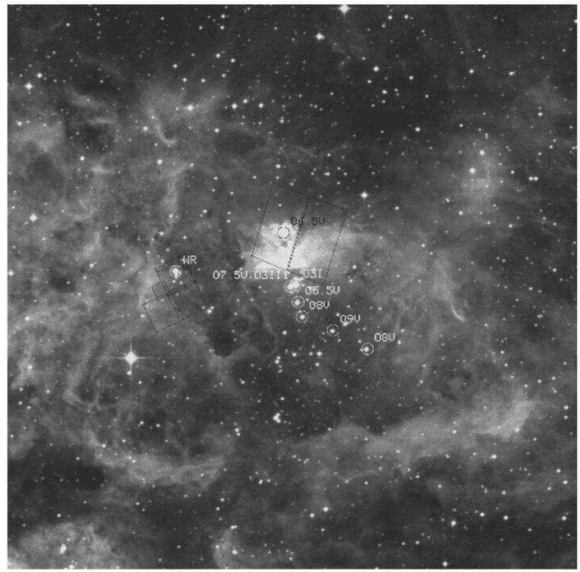}
\caption{A 
25x25~arcmin$^2$
DSS/POSSII (red filter) image of Pismis 24 with an overlay of the ACS and WFPC2
apertures used. North is toward the top, east toward the left. 
The O stars and one Wolf-Rayet star are marked (see Table~\ref{tab:stars}).
\label{fig:pismis24_apertures}}
\end{figure}

\begin{figure}
\plotone{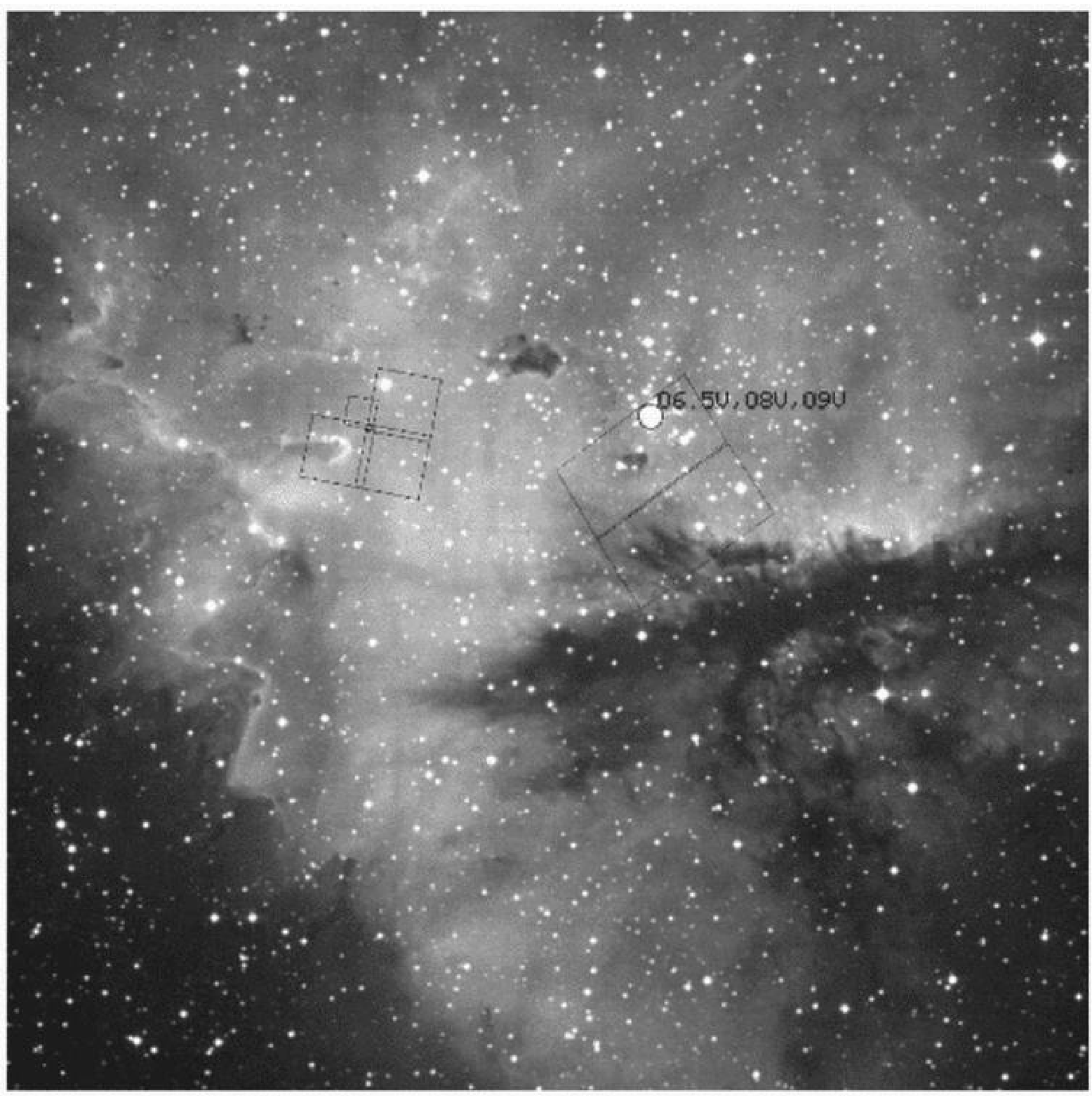}
\caption{A 25x25~arcmin$^2$
DSS/POSSII (red filter) image of IC~1590 with an overlay of the ACS and WFPC2
apertures used. North is toward the top, east toward the left. 
The O stars are marked (see Table~\ref{tab:stars}).
\label{fig:ic1590_apertures}}
\end{figure}

\begin{figure}
\plotone{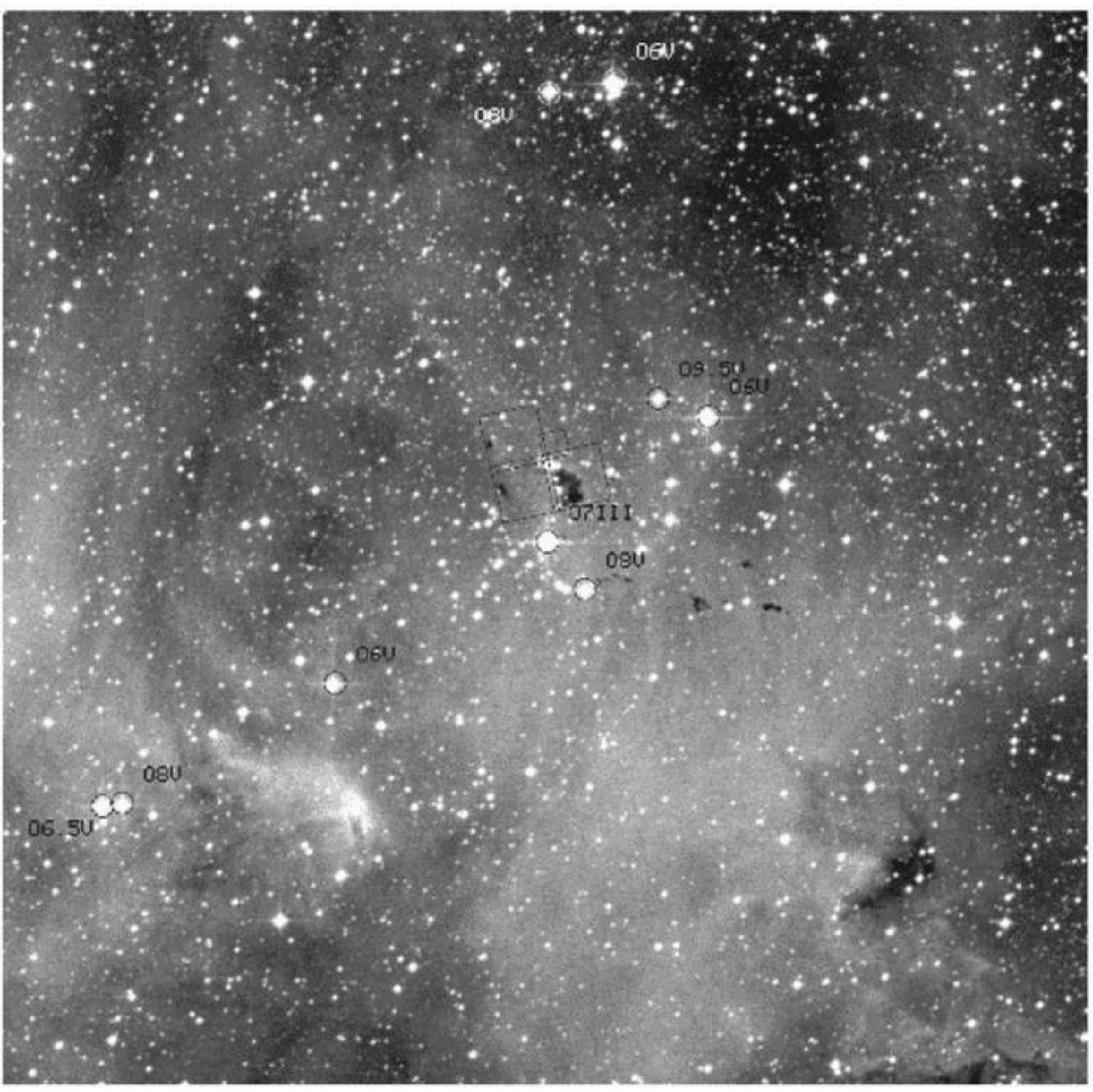}
\caption{A 25x25~arcmin$^2$  DSS/POSSII (red filter) image of IC~2944 with an overlay of the WFPC2
apertures used. North is toward the top, east toward the left. 
The O stars are marked (see Table~\ref{tab:stars}).
\label{fig:ic2944_apertures}}
\end{figure}

\begin{figure}
\plotone{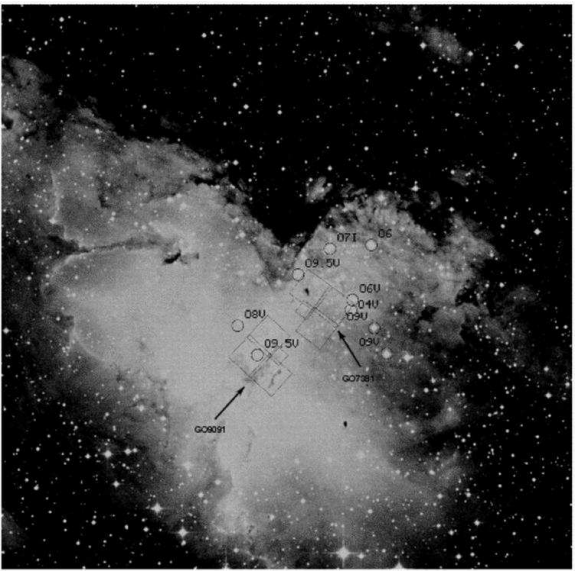}
\caption{A 25x25~arcmin$^2$ DSS/POSSII (red filter) image of M~16 with an overlay of the WFPC2
apertures used by programs GO9091 and GO7381. North is toward the top, east toward the left. 
The O stars are marked (see Table~\ref{tab:stars}).
\label{fig:ngc6611_apertures}}
\end{figure}

\begin{figure}
\plotone{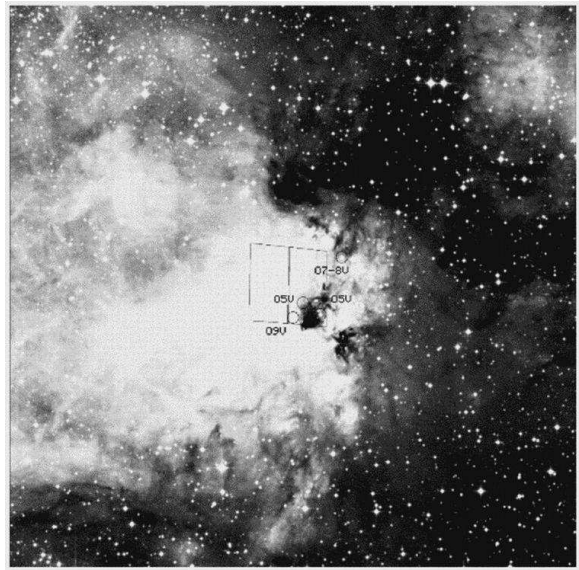}
\caption{A 25x25~arcmin$^2$ DSS/POSSII (red filter) image of M~17 with an overlay of the ACS
aperture used by program GO8992.
North is toward the top, east toward the left. 
The O stars are marked (see Table~\ref{tab:stars}).
\label{fig:m17_apertures}}
\end{figure}

\begin{figure}
\plotone{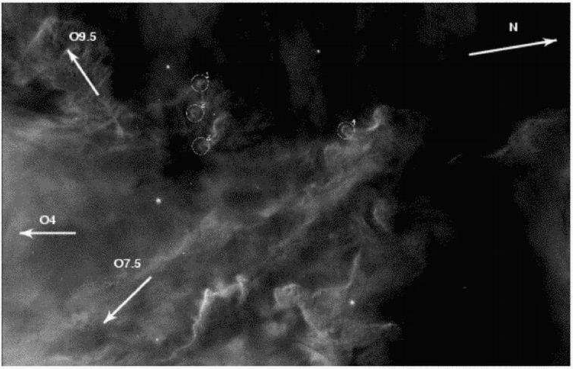}
\caption{A 100x65~arcsec$^2$ part of the ACS/H$\alpha$ image of NGC~6530 (Visit 2; 
pixel ranges x=2055-4075, y=530-1830 of the ACS chip).
North is indicated, east is left of north. The marked feature
numbers refer to the IDs in Table~\ref{tab:features}.
Approximate directions (arrows) to the main ionizing stars (labeled) are shown. For the 
exact location of the stars refer to Fig.~\ref{fig:ngc6530_apertures}.
\label{fig:visit2}}
\end{figure}

\begin{figure}
\plotone{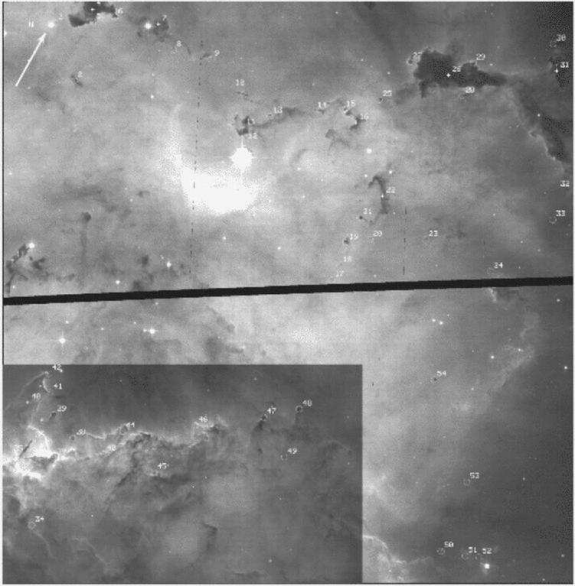}
\caption{The ACS/H$\alpha$ image of NGC~2467.
North is indicated, east is left of north. The marked feature
numbers refer to the IDs in Table~\ref{tab:features}. The main ionizing
star is labeled. The lower portion of the image is 
displayed with a larger greyscale stretch to provide
details in the brightest regions.
\label{fig:visit4}}
\end{figure}

NGC~6530 is a an extremely young open cluster in the Sgr OB1 association, located
on the eastern edge of the Lagoon Nebula (M~8). The associated \ion{H}{2} region has a roughly circular appearance with
diameter $\sim$18~pc (at a distance of 1.8~kpc; \citealp{Sun+00}). The walls of the cavity being excavated by
the ionizing sources are visible to the 
north and south
of the \ion{H}{2} region and appear fairly sculpted
with 
pillar-like
and other structures (Fig.~\ref{fig:ngc6530_apertures}). There are some dark shapes in the
DSS images with typical sizes of 0.015~pc, likely to be fragments of neutral material 
broken off 
from the walls by the UV radiation.

NGC~6530 contains three O stars immediately surrounding the ACS field of view (FOV; Table~\ref{tab:stars}).
They are Herschel 36, an O7.5~V star, 9~Sgr, an O4~V
star and HD164816, an O9.5 III-IV star. There is also an O6.5~V+O6.5 
binary (HD166052) about 19~arcmin (or $\sim$10~pc at 1.8~kpc) eastward of the main group.
Herschel 36 is thought to have a pre-main sequence
companion \citep{All86}.
The rest of the region (to the east of the O star cluster)
is dominated by B stars.  NGC~6530, and its population of pre-main sequence
stars, has been intensively investigated
by \citet{Sun+00}, who also determined the distance
to the cluster and its age (1.5~Myr with an age spread of 4~Myr).

In this region, \citet{Ste+98} have identified an ultra-compact \ion{H}{2} region
(G5.97-1.17), that they interpret as a circumstellar
disk around a B0 star, which is being externally photo-evaporated by Herschel 36.
The H$\alpha$ emission extends over 1080~AU, while the chord of the bow shock pointing toward 
Herschel 36 is $\sim$600~AU. This is possibly one of the 
structures thus far discovered most similar to the Orion Nebula proplyds, whose cusp chords range in size between
50 and 440~AU \citep{Bal+00}.
There are a few Orion Nebula proplyds with even larger sizes with
two cusps having chords of 1700~AU \citep{HO99,Bal+05}
at 430~pc \citep{WH77}.

We observed this region twice, the first time by aiming the ACS/WFC at a bright part of the region dominated by B stars
(Fig.~\ref{fig:ngc6530_apertures}).
The resulting ACS image appears bright and fairly smooth. Only on the 
eastern corner is there
more extinction.
No distinct features appear in this image and we do not discuss it further.
Our second visit image was taken between the three O stars. 
In Fig.~\ref{fig:visit2} we present a portion
of the ACS image, where numerous short ridges are visible. Some
of the ridges have protrusions and fingers pointing somewhere
between the O4~V and O9.5~III-IV stars. The look of this region is very different from that of the 
others
in that there are a lot of small bright features that give the region an ``embossed" appearance. 

There are 4 objects (\#1-4) that are reminiscent of proplyd cusps, although their base might be attached to a ridge. 
The smallest and best defined of the four is \#3. Its chord is 900 AU. The largest of them, \#4,
has a very rounded head and looks more like a protrusion from one of the faint ridges. 

\subsection{NGC~3324}
\label{ssec:ngc3324}

This \ion{H}{2} region is dominated by two O stars very close to one another, with spectral types
O6.5~V(n) and O8.5~Vp  (Table~\ref{tab:stars}), located in the middle of a
circular cavity with diameter 12~pc (at 3.0~kpc; \citealp{Car+01}; Fig.~\ref{fig:ngc3324_apertures}). 
They determined the possible presence
of pre-main sequence stars and an age for the region of 2-3~Myr (which however could be an upper limit
if the A0~I star
HD92207 is not a member of the cluster). The cavity walls, as seen in the DSS image, 
are remarkably smooth. This is also the case at the higher resolution of the WFPC2 image,
where only small protrusions can be noticed while the space immediately behind the ionization front is 
completely smooth with no features. The ACS image of this region, which included the two O stars,
is very smooth too, with only a few areas where the opacity is higher.
Given the lack of interesting 
features,
we do not present either the ACS or the WFPC2 images of this
region.
As we will discuss later (Section~\ref{ssec:ngc2467}), 
smooth ionization fronts do not appear to be associated with
fragments.

\subsection{NGC~2467}
\label{ssec:ngc2467}

This \ion{H}{2} region is about 8~pc (at 4.1~kpc; \citealt{FV89}) 
in diameter (Fig.~\ref{fig:ngc2467_apertures}) and
is dominated by one O6~Vn star (Table~\ref{tab:stars}).
There is also an O7 star 10~pc to the
east, but its radiation would be much weaker than that of the 
O6~Vn star.
The O6~Vn star is number 19 in the catalogue of \citet{Lod66} who determined the cluster distance to be
1.355~kpc. However, from 
an analysis of the colors of 71 stars in the NGC~2467 region, \citet{FV89} 
concluded that the cluster is actually the superposition of two clusters with the earlier type stars
being the farthest. This would position the 
O6~Vn
star and its associated \ion{H}{2} region (Sh~2-311) at a distance of 4.1 kpc.
The only age estimate of the cluster
was that of \citet[7~Myr]{Lod66},
which might be uncertain. 

Our ACS image FOV is centered in the middle of the cavity and it includes 
the
O6~Vn 
star (Fig.~\ref{fig:visit4}).
Despite the large distance to this \ion{H}{2} region, which makes it 
difficult to find any small-scale features (we estimated
that 3 pixels, equivalent to 610~AU, are a minimum to discover a feature),
the region is extremely rich in structure. 
Within the FOV we see
several extremely sculpted limb-brightened ridges and many small and partly limb-brightened fragments. 
There are two ridged regions: a cluster of layered,
structured ridges in the lower part of the ACS chip and
a fainter, sculpted ridge running down the right side of the chip.

The extremely sculpted ridge cluster
is very bright because of the diffuse emission from the ionization front. 
With no doubt, it is being sculpted by the
O6~Vn 
star, since the large attached fragment (\#40), as well as a number of thin fingers (e.g., \#36, 38, 39)
protrude toward it and are limb-brightened in that direction.
On the top of fragment \#40 there are
two fingers, \#41 and 42, with rounded tips and small, bright jets emanating from their tip and
base, respectively (see also Section~\ref{ssec:jets}). 
To the right of the bright ridge cluster, where the brightness subsides,
we see some cusps (\#50-53) which
appear to have once belonged to the ridge and might in fact have been the tips of thin fingers
such as \#24 or \#35, but have now been separated by the advance of the ionization front.

The ridge to the right is fainter (Fig~\ref{fig:visit4}). It is also better defined with a clear edge,
beyond which the area is uniformly dark. The reason for the difference between this ridge and the
bright layered ridges might be the 
viewing angle.
The bright layered ridges 
could be farther away from us than
the 
ionizing
star. This would mean that the whole ridge surface (i.e., the main
ionization front as well as what is behind the front) 
is illuminated from above. The faint ridge, on the other hand, could be
slightly closer to us than the ionizing source, such that its body (the area of molecular cloud 
on the other side of the ionization front) is not illuminated and appears
completely dark.

The faint ridge has a semi-detached        
large fragment (\#26; oddly, this fragment appears to be brightened 
{\it also} from the north-east, but there is no cataloged ionizing source in that direction;
the limb brightening could be the result of a complex projection effect). 
Similarly to fragment \#40, fragment \#26 has several protrusions (\#28, 29) and
even some smaller fragments (\#27) which are breaking off from it. Some 
additional fragments appear to have been left behind
by the advance of the ionization front (e.g., \#16 and \#22). 

These isolated fragments are close to the line of sight to the O6~Vn 
star. Most of them are dimly limb-brightened (e.g., \#13, \#15),
but some of them are not (e.g., \#25). These fragments are not dissimilar to those 
in IC~2944 analyzed by \citet{Rei+03}, but 
have
the following 
differences:
(i) These fragments are not embedded in a smooth
medium, but in a medium with variable patchy extinction. Most of them are limb-brightened, contrary
to those in IC~2944. (ii) The 
NGC~2467 fragments are less dark than those in IC~1590 and their
shapes include some very complex thin ones (e.g., \#10), while in IC~2944 most fragments are very dark and very stocky. 
(iii) Many of these fragments 
contain cusp-like
sub-units,
which appear to be about to separate (\#15 seems to be separating from 
\#16, but it could also be that the two fragments share similar lines of sight).
Some sub-units are seen nearby, but definitely separated (e.g., \#13) from the larger fragments. 
The smallest of these
protrusions (\#12), with a chord of only 2.5~arcsec, or 1025~AU at 4.1~kpc is
still attached to fragment \#11.
These rounded protrusions which are still attached or very close to the parent fragment are very similar in appearance to 
others that are definitely separated, such as \#32, \#33, \#50 or \#53. 
(iv) In NGC~2467 there
are no isolated circular fragments, as is the case in IC~2944. (v) Last, in NGC~2467 the ionization front is
relatively near the fragments; its proximity and appearance makes 
one
suspect that 
the fragments have been left behind by the advance of the front.
No such clear conclusion can be drawn for IC~2944.

\subsection{Pismis~24}
\label{ssec:pismis24}

The extremely young cluster Pismis~24 was studied recently by \citet{Mas+01} who determined
a high degree of co-evality for its numerous young O stars (0.7~Myr). 
As can be appreciated from Fig.~\ref{fig:pismis24_apertures},
this is a very large \ion{H}{2} region ($\sim$50~pc in diameter), although 
within it, there is a smaller and brighter \ion{H}{2} region, which is only
about 2.5~pc in diameter. The ACS and WFC chips 
were
placed on this region of intense emission
known as G353.2+00.9 (Nr. 1404 in \citealp{Pal+03}) to the 
south
of the main \ion{H}{2} region.
Only one O star is embedded in the \ion{H}{2} region, the others are grouped to the south.

\begin{figure}
\plotone{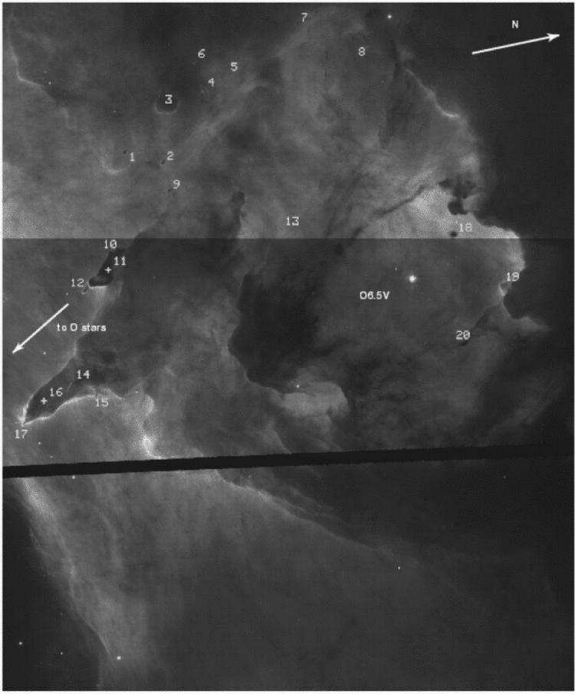}
\caption{A 110x75~arcsec$^2$ portion of the ACS/H$\alpha$ image of Pismis 24 
(pixel ranges x=229-2629, y=1135-4031 of the ACS chip).
North is indicated, east is left of north. The marked feature
numbers refer to the IDs in Table~\ref{tab:features}.
The approximate direction (arrow) to the main ionizing stars is shown,
as is one O star within the FOV. For the
exact location of the stars refer to Fig.~\ref{fig:pismis24_apertures}.
The upper and lower parts of the image are displayed with a different gray-scale stretch.
\label{fig:visit5}}
\end{figure}

\begin{figure}
\plotone{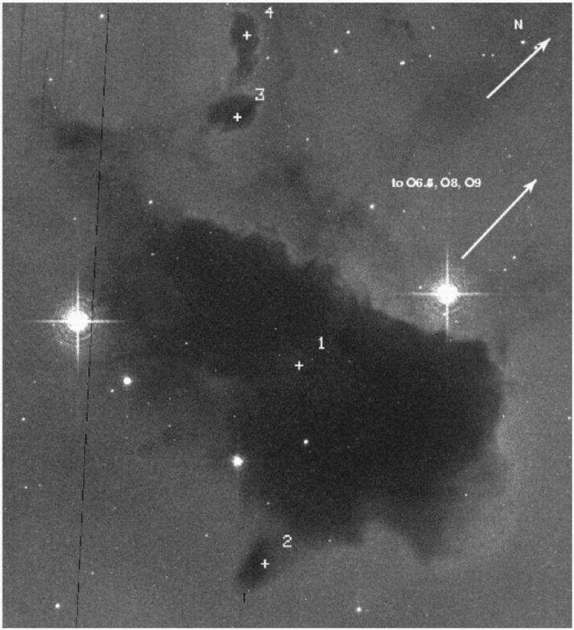}
\caption{A 50x55~arcsec$^2$ portion of the ACS/H$\alpha$ image of IC~1590 
(pixel ranges x=1390-2390, y=2760-3860 of the ACS chip).
North is indicated, east is left of north. 
The marked feature numbers refer to the IDs in Table~\ref{tab:features}.
Approximate directions (arrow) to the main ionizing stars (labeled) are shown. For the
exact location of the stars refer to Fig.~\ref{fig:ic1590_apertures}.
O stars within the FOV are labeled.
\label{fig:visit6}}
\end{figure}

\begin{figure}
\plotone{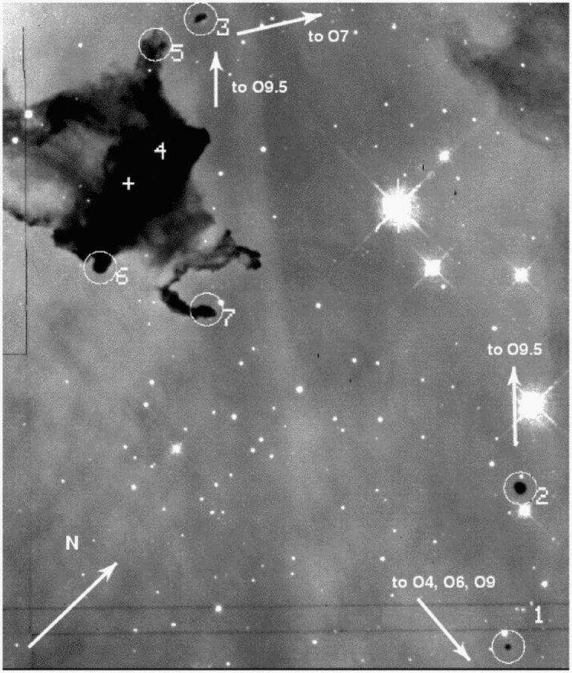}
\caption{A 35x40~arcsec$^2$ portion of the WFPC2/H$\alpha$ image of M~16
from program GO7381
(cropped to display the pixel ranges x=760-1460, y=710-1530).
North is indicated, east is left of north. 
The marked feature numbers refer to the IDs in Table~\ref{tab:features}.
Approximate directions (arrows) to the main ionizing stars (labeled) are shown. For the
exact location of the stars refer to Fig.~\ref{fig:ngc6611_apertures}.
\label{fig:ngc6611}}
\end{figure}

The ionizing stars are concentrated in two distinct locations.
The first and most important location is home to a very compact 
cluster of three O stars 
(O3~I, O3~II, and O7.5~V;
Table ~\ref{tab:stars}) just off the southern 
edge of the ACS FOV. At the second location we find one
isolated 
O6.5~V 
star located within the ACS FOV (Fig.~\ref{fig:visit5}).
We will refer to these locations as ionization sources
A and B, respectively. Off the edge of the ACS chip, in the same direction as
the stars in location A, but farther away, are four additional O stars, 
which are less tightly grouped and might or not contribute significantly
to the ionization of the features seen within the ACS FOV.

There is one long ridge extending from the top-left to the bottom-right of the ACS FOV 
(east to west; Fig.~\ref{fig:visit5}),
which changes character as it runs along, from very sculpted to smooth. 
This ridge has one large and one smaller protrusion (\#16 and \#11), both similar to the
large attached fragment in NGC~2467 (\#40 there). Both protrusions end with a thin finger (\#17 and \#12).
This ridge appears to be formed and ionized by photons from source A.
A second ridge is layered on it (top-right corner in Fig.~\ref{fig:visit5}), 
but clearly limb-brightened by the star at location B.
Facing the first ridge there is a third, smooth ridge (seen cutting the bottom left corner
of Fig.~\ref{fig:visit5}), which appears to be limb-brightened from the 
direction of the O6.5~V star in source B. 

A group of several cusps are observed in this region (top-left corner of Fig.~\ref{fig:visit5}), 
some with peculiar shapes. 
Typically
all chord sizes are close to 2000~AU (at 2.5~kpc)
although the smallest cusp (\#8) has a chord length of 500~AU. All of the cusps are
between the first and second ridge.
Some of them (\#1, \#2, \#7, \#8 and \#9) point toward ionizing location A, while others
(\#5 and \#6) point somewhere between A and B.  Another cusp (\#4) 
in the same group is a very long (7700~AU) bacterium-shaped object,
pointing toward source B, with both sides of its long body 
limb-brightened.
 A second source
 (\#3) is larger (12\,000x15\,000~AU) and
has the shape of an upside down heart, with very smooth, limb-brightened edges
which again 
appear
to be illuminated from both sides. We call this a {\it cusp} because of its 
smooth, limb-brightened 
edges and its location close to other cusps, but we note that its size is closer to
a {\it fragment} than a cusp.

Near the second ridge we see one of the strangest objects yet encountered: a circular 
object (\#18), seen in silhouette and not limb-brightened, with a dark oval core of
size 2625x2125~AU, embedded, but not centered, in an oval, grey halo 5380x4000~AU.

Finally, the ionized area between the two smooth ridges (between the lower part of ridge one
and
ridge three;
lower left corner of Fig.~\ref{fig:visit5}) is fairly smooth and devoid of features, reinforcing the
conclusion already drawn for NGC~3324
(Section~\ref{ssec:ngc3324})
that fragments and cusps are seen only in association with very sculpted ridges.

\subsection{IC~1590}
\label{ssec:ic1590}

IC~1590 is an \ion{H}{2} region with an approximate size of 21~pc at 
2.94~kpc
(Guetter \& Turner 1997). 
On the eastern side of the DSS image
(Fig.~\ref{fig:ic1590_apertures}) we find the ionization front eating into the molecular cloud. Several
protrusions are seen extending out from this ridge. Our WFPC2 image is centered on one such trunk,
but the low SNR prevented us from being able to study it in detail. 
The stellar content of this region consists of a rich and compact cluster 
comprising a few O stars with spectral types between O6.5~V
 and O9~V,  possibly all in binary and even triple systems
(Guetter \& Turner 1997; Table~\ref{tab:stars}). 

Dark fragments can be seen
in the DSS image scattered around the ionized region. Within the FOV of the ACS chip, which includes the
ionizing sources, we see only one such fragment (\#1),
very similar to the ``Thackeray globules" \citep{Tha50}
studied by Reipurth et al. (2003). Unfortunately this image is underexposed 
and we can therefore only distinguish two smaller fragments (\#3 and \#4) that appear to have broken off
from the main fragment. These two fragments have sizes of 8500~AU and 6800x20\,000~AU,
respectively and are therefore comparable to 
fragments breaking off the ridge in NGC~2467 (e.g., \#27 there).

Although these fragments are not very close to the eastern ionization front, they are close to an
area of high extinction covering the south corner of the ACS chip. This extinction is likely due to 
neutral material in the foreground on the near side of the region. Whether the fragment on the ACS chip
has broken off from the ridge on the east side of the DSS image, or from another, hidden ionization 
front is
hard to say. We note that the presence of a large fragment seen in silhouette against the ionizing sources
and not obviously close to a ionization front, is similar to the situation encountered in
IC~2944 \citep{Rei+03} and M~16 (Section~\ref{ssec:ngc6611}).

\subsection{IC~2944}
\label{ssec:ic2944}

IC~2944 is a large faint \ion{H}{2} region about 28~pc in diameter 
(at a distance of 1.8~kpc; \citealt{AM80}). 
It is ionized by a cluster of nine O6-O9.5~V stars
which was studied by \citet{Wal87}, who concluded that the stars belong to one cluster, rather than
being a chance superposition of more than one cluster. He determined a distance to the cluster
of 2.0~kpc based on the stellar types. The cluster's age is 7~Myr \citep{Dut+03,Bic+03}.
This is the oldest cluster in our sample.

\citet{Rei+97} studied the ``Thackeray's globules" in this region from ground-based
imaging. 
They
adopted a distance of 1.8~kpc, previously obtained by Ardeberg \& Maurice (1980)
from nebular kinematic arguments. Given the similarity of the two distance estimates and the fact
that we will make extensive comparison with the study of Reipurth et al. (1997), we will
adopt their value. From the lack of CO emission,
Reipurth et al. (1997)
concluded that the globules are not star-forming. 
They are instead transient fragments
of neutral gas and dust that broke off from the main walls. The dynamical lifetimes of the
clumps,
of the order of 1~Myr, obtained from kinematic arguments, 
agree with transient clump lifetimes predicted from magnetized hydrodynamical models of \citet{Vaz+05}.
(These models do not include the effects of radiation, which is likely to be an important factor in the
destruction of the clumps. However, we know of no model predictions which does include the effects of radiation,
so this is as good a comparison with theory as we can make at present, although one could suggest that the
model timescales are only an upper limit).

We have re-analyzed the WFPC2 image of Reipurth et. al (2003). In Table~\ref{tab:features} we list
the fragments using their numbering system; we precede their numbers 
by an ``R" to
separate those fragments identified by them from a few that they did not list. We do not present
the WFPC2 image, since it can be seen in that paper.
The fragments are seen silhouetted (and sometime dimly limb-brightened) 
against a bright uniform background.
The O stars in the region
are scattered to the north, north-west and south of the FOV. The closest star is an 
O7~III 
star just south of the FOV.
The next closest star is an 
O8~V 
star 0.7~pc to the south-west. All the other stars are farther (Fig.~\ref{fig:ic2944_apertures}).
The fragments
range in shape from 
complex to perfectly circular.  The largest fragment (\#R1; 30\,000x50\,000~AU) is very complex and 
structured, with dozens of appendages. This can be compared to fragments \#1 in IC~1590 and
\#4 in M~16. 
Other fragments are smaller but still have irregular shapes (\#R2, \#R30/1;
18\,000x10\,000~AU and 6000x3200~AU, respectively; Fig.~\ref{fig:proplydstamps}). 
Smaller fragments have elongated or rounded, though slightly
asymmetric shapes (e.g., \#R37 and \#R32; 1500x3200~AU and 2200~AU, respectively; 
Fig.~\ref{fig:proplydstamps}). 
Finally, the smallest fragments
are either perfectly circular (\#R33 or \#R27; 1800~AU and 1000~AU, 
respectively), or too small to resolve their shape
(e.g., \#R40A, B and C [Fig.~\ref{fig:proplydstamps}]
and \#R41; the first three have a diameter of $<$500~AU, while the
last is 540~AU wide).

\subsection{M~16}
\label{ssec:ngc6611}

The stellar content of M~16 (also known as NGC~6611 or the Eagle Nebula)
was investigated by \citet{Bel+99} and \citet{Bel+00} who determined the
distance to the cluster to be 2.14~kpc and its age to be 6~Myr with a
similar age spread. 
The bright \ion{H}{2}
region has an approximate diameter of 18~pc and it is very rich in O stars: 
at least nine O stars with spectral types
between O4 and O9.5 are found in the region (Table~\ref{tab:stars}).

M~16 was thoroughly investigated by Hester et al. (1996). They
observed the region near the large dusty ``elephant trunks" (a set of three dusty pillars; their FOV is marked in 
Fig.~\ref{fig:ngc6611_apertures}) and
concluded
that the several cusps near the ``elephant trunks" 
are star-forming nodules left behind by the ionization front eating
into the molecular cloud. We have looked at a WFPC2 archival image of a slightly different part of the
region. Our FOV (placed only a few arcminutes to the north-west;
Fig.~\ref{fig:ngc6611}) 
contains a large fragment (\#4) similar to a typical 
``Thackeray globule"
with a very long, round-tipped protrusion (\#7), and two perfectly circular fragments (\#1 and \#2;
750~AU and 1500~AU in diameter, respectively; Figs.~\ref{fig:ngc6611} and \ref{fig:proplydstamps}) similar to those encountered in IC~2944.

\subsection{M~17}
\label{ssec:m17}

M~17 is one of the brightest \ion{H}{2} regions in the galaxy. \citet{HHC97} studied its massive star content 
and determined the presence of a large number of massive YSOs and an age for the cluster of only 1~Myr. 
They also determined the distance to the region to be 1.3~kpc.

\begin{figure}
\plotone{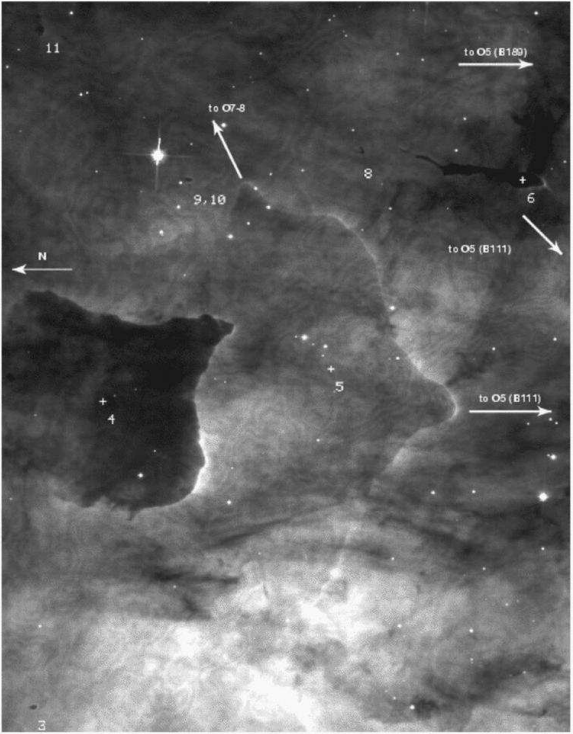}
\caption{\footnotesize{An 80x60~arcsec$^2$ portion of the ACS image of M~17
(image j8cw07041\_drz.fits, rotated by 90 degrees, and cropped to display the pixel ranges x=444-1716, y=2558-4206).
North is indicated, east is left of north. 
The marked feature numbers refer to the IDs in Table~\ref{tab:features}.
Approximate directions (arrows) to the main ionizing stars (labeled) are shown. For the
exact location of the stars refer to Fig.~\ref{fig:m17_apertures}.}
\label{fig:m17}}
\end{figure}

\begin{figure}
\plotone{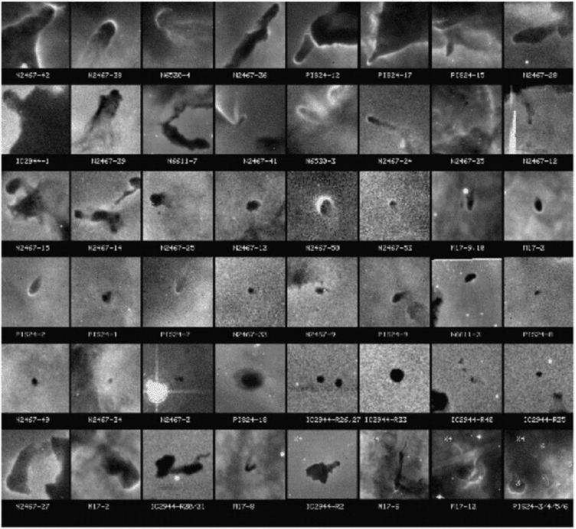}
\caption{Subset of the features identified in the \ion{H}{2} regions. 
Each stamp is 5~arcsec on a side,
except for the last four ``stamps" which are 20~arcsec (marked x4).
Stamp labels refer to the region within which the features are found and the IDs from 
Table~\ref{tab:features}. For an explanation of the layout see Sec.~\ref{sec:thestarclusters}.
\label{fig:proplydstamps}}
\end{figure}

The region within the ACS FOV is very rich, with diffuse features not seen in the other images. In Fig.~\ref{fig:m17}
we show a portion of the FOV with two large, but oddly very smooth, 
fragments (\#4 and \#5) which might be attached to a ridge
outside the FOV. There are also three classical cusps in this image 
(\#3, \#9 and \#10), with
chords of 720, 320 and 650~AU, respectively.
Cusp \#9 
is the smallest
cusp detected outside the Orion Nebula and is close to the 
3-pixel
detection limit
for a feature in this region. 
There is also a most curious long fragment (\#6)
with a limb-brightened cusp facing one of the hottest ionizing sources. 
From it protrudes a bright straight jet
(see also Fig.~\ref{fig:jets} and Section~\ref{ssec:jets}). The other side of the
fragment consists of a very thin tail. Just north of the tail is a 
small
boomerang-shaped
fragment seen in silhouette (\#8).
There is one circular silhouette in this region (\#7; not included within the FOV of
Fig.~\ref{fig:m17} but shown in Fig.~\ref{fig:m17_7}) in the middle of an 
area of extremely bright emission. 
It also appears to have a halo, 
similar
to 
silhouette \#18 in Pismis~24. The inner, darkest part of the feature 
is shaped like a boomerang (Fig.~\ref{fig:m17_7}).

\begin{figure}
\plotone{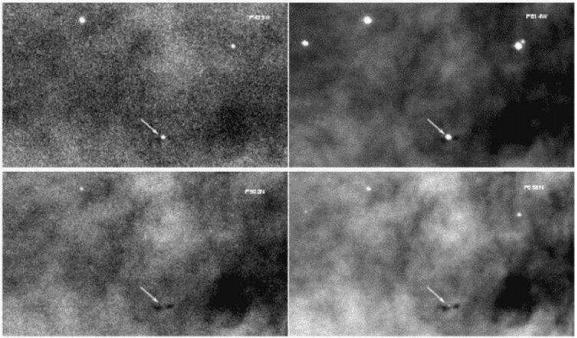}
\caption{A 16x11~arcsec$^2$ portion of the ACS image of M~17 in the immediate area
of M~17-12  (Table~\ref{tab:features})
which is located at pixel position (2022,3844),
imaged through four different filters
(labeled).
North is toward the right, east is up. The arrows point to the
putative star with disk.
\label{fig:m17_proplyd}}
\end{figure}

\begin{figure}
\plotone{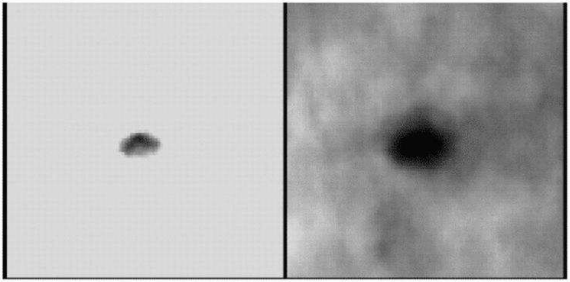}
\caption{A 5x5~arcsec$^2$ portion of the ACS/H$\alpha$ 
image of M~17, centered around feature 7.
Two greyscale levels are used to show both the boomerang 
shape of the darkest absorption (left panel) and
the rounded shape of the overall feature (right panel).
\label{fig:m17_7}}
\end{figure}

\begin{figure}            
\plotone{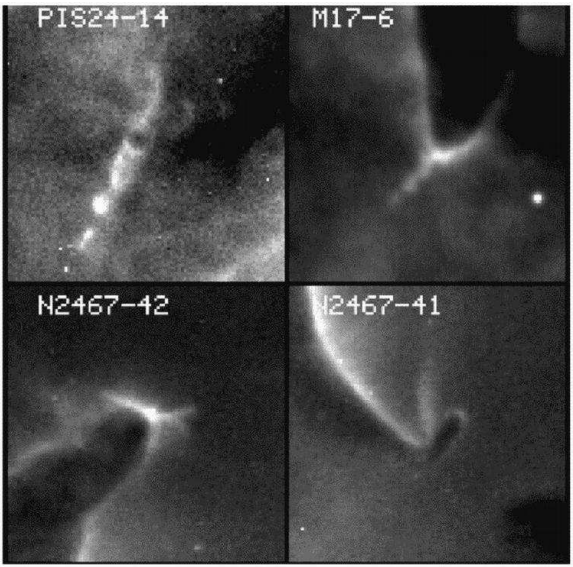}
\caption{5x5~arcsec~$^2$ images of the four jets observed in the \ion{H}{2} regions. 
North is oriented as in Figs.\ref{fig:visit5} 
(Pismis~24), \ref{fig:m17} (M~17), and 
\ref{fig:visit4} (NGC~2467). Labels refer to the regions
as well as the ID numbers from Table~\ref{tab:features}. 
The jets are discussed in Section~\ref{ssec:jets}.
\label{fig:jets}}
\end{figure}

In M~17, we also find the only fragment (\#12; not included within the FOV of
Fig.~\ref{fig:m17}) with a possible central star. 
This silhouette looks like an edge-on proplyd disk.
In Fig.~\ref{fig:m17_proplyd}, we show four 
images of the same region of sky in four different filters. Around 
one of the stars, we see what could be 
an edge-on
dusty disk, 1170~AU in diameter and 
195~AU
thick. The star at its center is a faint red star. The object most similar
to this, is Orion 114-426, where an 860-AU edge-on disk partly obscures a central
star \citep{MO96}. We notice that the central star in our fragment is not exactly centered. This however could be similar
to the situation encountered in HH513 (Orion 165-235) where the star embedded in the disk is offset from the disk's center, 
and it is thought to be the companion of a hard-to-detect central star 
\citep{Bal+00}. An alternative explanation was pointed out by
the Referee: what is seen as the central star could also be reflection of stellar light
from the inner rim of the disk. This would appear off-center if the disk's inclination angle
was not exactly 90~degrees.

\section{Discussion}
\label{sec:analysis}

Below, we group some of the most interesting characteristics of the features 
found in the \ion{H}{2} regions and compare
them across regions and with results from the literature. 

\subsection{The shape, size and limb-brightening properties\\
 of the proplyd-like fragments}
\label{ssec:SSLBproperties}

With the observation of a large number of \ion{H}{2} regions one could hope to find proplyd-like fragments
bridging the gap between the {\it bona fide} Orion Nebula proplyds (where a star is surrounded by a
dark disk seen in silhouette, often embedded in a bright, tear-shaped
cusp) and the bright, similarly tear-shaped cusps with no central stars, found 
in other regions. 

In order to compare the fragment sizes and size 
distributions 
found in different regions,
we have converted the angular sizes to linear sizes (in
AU) using the distance estimates from Table~\ref{tab:clusters} 
and calculated the effective radius of each fragment. 
The effective radius is defined by \citet{Rei+03} as the square root of the fragment's area. 
In order to compare with their results we have maintained the same definition.
If the fragment is rounded and is characterized by only one
measurement in Table~\ref{tab:features}, we assume that fragment to be a circle and
the measurement to be the diameter. 
If the fragment has a short and a long measurement
in Table~\ref{tab:features}, we assume it is a rectangle. In this way some areas are overestimated,
in particular for intermediate size, complex fragments such as some of those
in NGC~2467. However, the smallest fragments, which are almost always rounded,
are well represented by our calculation. Overall,
our approximation has no bearing on the conclusions.

\citet{Rei+03} carried out a similar analysis and concluded that if
{\it photo-evaporation} was the only process determining the fragment size distribution,
there should be fewer small fragments.
They therefore suggested that {\it fragmentation} produces 
small fragments at a high rate. Their size distribution has two peaks, one at less than 1~arcsec ($<$1800~AU) and one
at 1.5~arcsec (2700~AU). We reproduced their histogram using our measurements (Fig.~\ref{fig:sizes_all};
second-to-last panel). There, we see
a distribution with a broad peak centered at $\sim$1000~AU, extending between 
300 and 5000~AU.
We therefore
confirm the presence of a large number of objects smaller than 1800~AU but do not resolve the peak
at size 2700~AU (possibly because of our coarser 
measuring method).

NGC~2467 has a distribution peaked at larger sizes than IC~2944 with two peaks at sizes 
3200 and 8000~AU (or one
peak at $\sim$5000~AU). The smallest detected size (800~AU) is larger than in IC~2944, 
probably because NGC~2467 is much farther (4.1 vs. 1.3~kpc; 3 pixels = $\sim$600~AU). 
Pismis~24 has fewer fragments and the smallest fragments,
$\sim$790~AU across, are quite a bit larger than the 3-pixel detectability limit at 
2.5~kpc ($\sim$370~AU).  The largest fragments in Pismis~24 (which are cusps)
are similar in size to those in IC~2944 (which are silhouettes).
Finally, although we cannot assess the fragment size distributions for the other \ion{H}{2} regions,
we note that M~17 has fragments as small as those in IC~2944,
although they are cusps rather than silhouettes. We also note that
cusp \#9, found very close to cusp \#10 in M~17 
(M~17-9 in Fig.~\ref{fig:proplydstamps}), with a chord size of only $\sim$320~AU, is the smallest cusp ever detected
outside the Orion Nebula.

Proplyd cusps observed within the Orion Nebula
are the bright ionization fronts where the external radiation field ionizes the material
photo-evaporated from the circumstellar disks. Cusps found elsewhere, in particular where no star is present,
might be lumps of dusty gas externally photoionized by the nearby O stars, 
either on the way to forming a star or just transient. 

In Table~\ref{tab:prop-summary}, we summarize the size ranges of bright cusps and dark silhouettes 
from the literature alongside those determined here.
For the Orion Nebula, we have taken as 
representative
the proplyds in Fig.~7 of Bally et al. (2000).
The smallest bright cusp found there, 179-353, has a chord of $\sim$50~AU. The largest,
206-446, has a chord of 440~AU. However, Orion Nebula cusp chords can be as large as 1700~AU, as are those
of proplyds 244-440 \citep{HO99} and 181-826 \citep{Bal+05}.
For the Orion Nebula silhouette disks the range is 65-800~AU (where at the small
end we find 182-332, while at the large end we find 114-426). We use these 
values as fiducial marks
for our comparison.

Outside the Orion Nebula, the only objects 
that can be classified 
as possibly {\it bona fide} disks,
because of the presence of a central star,
are the disk found in M~20 (Table~\ref{tab:prop-summary}; \citealt{Yus+05}),
the cusp found in M~8 (Table~\ref{tab:prop-summary}; \citealt{Ste+98}),
some cusps with central stars found in M~16 
(Table~\ref{tab:prop-summary}; \citealt{Hes+96}),
as well as the edge-on disk found by us
in M~17 (but see section \ref{ssec:m17} for our reservations). 
The dark, perfectly circular silhouettes found in M~16 and IC~2944,
and the circular silhouette with a halo found in Pismis~24 
do not contain a central star. 

Some of the cusps in Pismis~24 and M~17 have sizes small enough (750~AU and 315-845~AU, respectively) 
to compare with some of the cusps in the Orion Nebula, but have
no central stars. 
As such, until the central star is detected,  
they remain in the same class as the larger (chords $>$800~AU)
starless cusps, although their sizes can be
similar to the sizes of the Orion Nebula proplyds and, as such, they
are closer to bona fide proplyds than, for instance, the fragments found in Carina
\citep{Smi+03}.
The large cusps with no central star found in Carina by \citet{Smi+03} and in NGC~6303 by
\citet{Bra+00} are likely to be fragments of molecular cloud. This is also likely to be the
case for the 
large-to-intermediate size cusps found in NGC~2467, although,
due to the higher resolution of the NGC~2467 observations, we can detect cusps with smaller 
sizes
than those detected by \citet{Smi+03} and \citet{Bra+00}.

Finally, we compare
the starless circular silhouettes found in IC~2944 and 
M~16 with
the tear-shaped cusps found within this work and elsewhere. 
These cusps and silhouettes can have similar sizes. We therefore
suggest that they are all self-gravitating fragments of molecular cloud:
they acquire a tear-drop shape and a photo-ionized limb
when embedded in the radiation field, but lose the bright limb and acquire a
more spherical shape when shielded from the direct radiation
(this is confirmed by models by \citealt{HO99}). 

The more important question, in the absence of a central star, is whether star formation
is taking place in these cusps and silhouettes.

\subsection{Are the circular silhouettes and the tear-shaped cusps star-forming?}
\label{ssec:simon}

The circular appearance of a number of silhouette fragments and the tear shape of cusps
in our sample could be easily explained if these objects are self-gravitating 
and are in the process of collapsing to form stars. 
This is also an implicit assumption of the claim made in Section 
\ref{ssec:SSLBproperties} that circular silhouettes acquire
their spherical shape when shielded from the direct ionizing
radiation which would otherwise shape them into tear-drops. 
\citet{Rei+03}, however, measured a rather low optical depth 
across the face
of some of their circular silhouettes, and deduced that they are not star forming. The jury appears
therefore to be still out on the star-forming properties of these fragments and their 
relationship to {\it bona fide} proplyds.

Using the same optical depth argument, we
compared the characteristic sizes of the circular silhouette fragments in IC~2944 and
M~16 with the gravitational
stability length scale, or Jeans length \citep{Jea1902}, for gas with the same 
density and temperature as the globules. 
Using the H$\alpha$ optical depth determined by \citet{Rei+03} ($\tau \sim 0.1$--1), 
we constrain the globules' total hydrogen column density along the line of sight to be 
$N_{\rm H} = 10^{20}$--$10^{21} \: {\rm cm^{-2}}$.
If we assume that the characteristic size of a given globule along the line of
sight, $L$, 
is similar to its observed size in the plane of the sky, or in other 
words that our circular globules are indeed approximately spherical, then we
can use this to constrain the mean volume density of the gas:
\begin{equation}
 \bar{n}_{\rm gl} \simeq 6.7 \times 10^{3} \left(\frac{\tau}{0.1}\right) 
 \left(\frac{L}{1000 \: {\rm AU}}\right)^{-1} \: {\rm cm^{-3}}.
\end{equation}
If the density distribution 
in
a typical globule is approximately uniform, then the
Jeans length within it will be:
\begin{equation}
 L_{\rm J} = 8.3 \times 10^{4} \left(\frac{T}{10 \: {\rm K}}\right)^{1/2} 
 \left(\frac{\tau}{0.1}\right)^{-1/2} \left(\frac{L}{1000 \: {\rm AU}}\right)^{1/2} 
 \: {\rm AU} \label{eq:LJ},
\end{equation}
where $T$ is the gas temperature.
Even for $\tau = 1$, this gives a value for $L_{\rm J}$ that is at least an
order of magnitude larger than the size of a typical spherical globule. Also,
while it would be possible in principle to reduce $L_{\rm J}$ by reducing the
gas temperature $T$, it does not appear plausible to have $T \ll 10 \: {\rm K}$;
instead, we would actually expect that $T > 10 \: {\rm K}$, due to photoelectric 
heating of the gas by the diffuse nebular emission. This line of reasoning would therefore 
lead to the same conclusion of
\citet{Rei+03}, namely 
that the small spherical silhouettes seen in IC~2944 and M~16 are not gravitationally bound.

This argument may contain a basic flaw because of an assumption made. The implicit assumption in both
the \citet{Rei+03} and the Jeans length arguments is that the globules have 
relatively flat internal density distributions. This assumption is made because of the
flat brightness distribution within the fragments.
It is however highly unlikely that the densities
within the fragments are constant:
unless the objects have the same physical thickness
across their faces, then a constant optical depth would mean that (for a
spherical object) the spatial density decreases towards the middle
owing to the path-length increasing. The most likely explanation is that the
optical depth flattening out at about unity (and being limited to about
unity) is due to filling-in of the light by the finite diffraction pattern of the
telescope-camera. The images of the fragments are surrounded by very bright
nebulosity and this tends to fill-in the central regions, thus limiting the
{\it apparent} optical depth and leading to underestimates of the spatial
density \citep{MO96}.
If instead the globules have strongly peaked central density 
distributions, then their inner regions will be more strongly bound than this simple 
argument suggests and indeed might be in the process of making stars. 

Besides their shapes, there is another argument in favor 
of the circular 
silhouettes and cusps being in the process of making stars.
Cusps are found only in proximity to the cavity walls,
indicating that they might have been recently freed from the molecular cloud walls by
the advance of the ionization front and 
indicating in turn that they are considerably denser than the 
molecular cloud mix. 
Based on this we would argue, although cannot prove, that these
tear-shaped cusps might well be self gravitating and star-forming.

This conclusion is also in line with the results of \citet{Smi+05b} who found the proplyds in the Orion Nebula 
only in close proximity to the youngest hot stars,
while farther from 
these stars
they only detect the proplyd descendants, mid-IR emitters interpreted
as stars with feebler disks.
In this scenario, stars of all masses are forming in the molecular cloud. 
What we 
see, however,
is dictated by the early evolution of the most massive of them. True proplyds
should be common, {\it but} they are only seen if they are illuminated by a nearby, very hot star,
{\it even if} they do not last long 
in close proximity to such intense radiation. 
When the proplyd disk has been sufficiently evaporated by the hot stars, 
the proplyd appearance vanishes
(it loses its bright skin), although its central star 
still retains the thinned out disk and is detected
as a mid-IR emitter, near what has become a slightly older O star.
In parallel with the emergence and demise of proplyds, the O star ionization front
progressively frees star-forming clumps, ionizing
their outer parts, shaping them into tear-drops and initiating their evaporation. As a result,
these cusp-like objects do not last
long, which 
explains 
why we see them only near the cavity walls, where they have been recently freed. 
The star(s) made in these clumps eventually emerge from their placental material,
but might never be seen as proplyds because they are
not sufficiently close to the ionizing source(s). 

With hindsight, the Referee's suggestion that we should have
observed \ion{H}{2} regions containing groups of pre-main sequence low-mass stars
in close proximity to O stars, is right. However, this further
constraint would have resulted in overall larger distances to our \ion{H}{2} region
sample, making the detection of small features harder still.

\subsection{Smooth ionization fronts}

In some regions, namely NGC~3324 and parts of Pismis~24 the ionization front is smooth. This means
that there are no protrusions
or broken
fragments of any size. 
Is this lack of features associated 
with
the type and age 
of ionizing stars and their geometric distribution?
Or is it a characteristic of the molecular cloud itself?
It looks like some molecular 
clouds,
for some reason, 
neither
fragment nor 
contain
small clumps waiting in the
molecular cloud to be freed by the passage of the ionization front.
This characteristic does not appear to be related to age since although
NGC~3324 is 
young,
so is M~17 
which has
a variety of features. Also, Pismis~24
has both sculpted (with nearby fragments) and smooth (with no nearby fragments) walls, 
confirming that the smoothness of the wall has nothing to do with the age of the region.
It appears more likely at this stage that the 
lack of fragments is due to the nature of the molecular cloud itself.

\subsection{Jets}
\label{ssec:jets}

Four jet-like structures have been observed in the \ion{H}{2} regions 
(Fig.~\ref{fig:jets}). Two of them (\#14 in Pismis~24 and \#6 in M~17) are
knotty and straight; \#14 is superimposed
on the large attached fragment \#16 and does not appear to be emerging from
any structure in particular. It is 8740~AU long and 1433~AU wide. 
It protrudes about 30$^{\rm o}$
to the south of the direction to the
ionizing stars. Fragment
\#6 in M~17 protrudes away from the cusp-like 
tip of a peculiar fragment (which we also labeled \#6; Fig.\ref{fig:proplydstamps}) and directly toward one of the ionizing stars. 
It is 1490~AU long and 390~AU wide. 
The other two jets are in NGC~2467,
projecting from two attached cusps (\#41 and \#42) which in turn protrude from the large attached fragment \#40.
Both of these jets are fuzzier and smoother (rather than knotty) than the first two, and they appear to bend. 
Jet \#42 has two ends, like the forked tongue
of a snake. The double jet originates from a bright ``knot" embedded in the bright limb of 
the cusp.
It is possible that only the upper tip is the true jet, while the lower one is
a chance superposition with a bright part of the region. The upper part measures 3275~AU from base to tip
and has a width of 410~AU. It protrudes in a direction about 
45$^{\rm o}$
west of the direction to the ionizing star.
The lower part is similar in length although it bends away from the ionizing source.
The second fuzzy jet, \#41, is peculiar in that it 
originates from the base, rather than the 
cusp of fragment \#41.
It measures 5730 x 1430~AU and
protrudes at about 
45$^{\rm o}$
east of the direction to the ionizing star. 
All four jets are brighter in emission line luminosity (F658N) than in continuum light (F550W).

A couple of dozen 
mono- and bi-polar 
micro-jets
are found protruding from the
Orion Nebula proplyds. These are smaller and less energetic than Herbig-Haro (HH)
jets, which are usually found only in neutral regions. 
The characteristic length of the 
micro-jets
is between 500 and 1000~AU, while
their widths range between 20 and 200~AU 
\citep{Bal+00}. 
Some of them are low ionization and are found within the ionization front of the proplyd,
while others extend beyond the cusp and have higher ionization levels. 
Most of these 
micro-jets
are mono-polar (for a discussion of this characteristic we refer the reader to section 5.4 of
\citet{Hen+02}). Most of the jets
visible in the Orion nebula are micro-jets, but there are also some larger HH objects.
These are seen as a series of aligned bow shocks which can extend 15\,000~AU.

Our jet \#14 in Pismis~24 appears intermediate between a 
micro-jet
and a HH object. Its knotty
structure could be interpreted as a series of bow shocks. Jet \#6 in 
M~17
has a size
comparable to the Orion Nebula
micro-jets.
The two jets in NGC~2467 have sizes that 
classify them as large micro-jets. However, 
several of their properties, such as their smooth and fuzzy appearance,
do not conform with the canonical jet characteristics.

\section{Summary}
\label{sec:summary}

Here we present a summary of our findings.

\begin{figure}            
\vspace{16cm}
\includegraphics{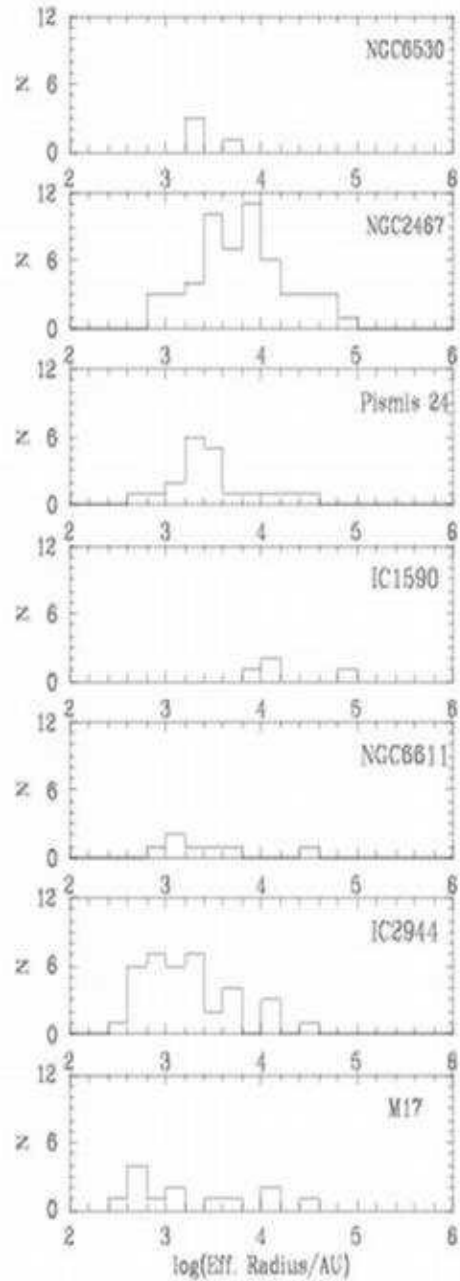}
\caption{A histogram of the $\log$(Effective Radius) of fragments found in each region.
Those regions with insufficient number of fragments to determine the distribution are
displayed to compare the sizes of the fragment with those of the other regions. 
\label{fig:sizes_all}}
\end{figure}

\begin{itemize}
\item The interface between ionized gas and neutral molecular cloud in \ion{H}{2}
regions can be smooth and devoid of fragments or sculpted and rich in fragments.
This points to the fact that even the small cusps, 
quite
similar to the proplyd cusps in
the Orion Nebula, are actually fragments of molecular cloud that have recently broken off and are
being photo-evaporated.

\item 
Four bright (micro-) jet candidates might point to on-going star formation.
However, none of them appear to originate from a star. Unless the absence of
the star can be explained by extinction, this characteristic makes their nature
suspicious. Central star aside, two of the jet candidates have all the
characteristics of micro-jets, while the other two appear smoother and fuzzier
than typical micro-jets observed in the Orion Nebula.

\item In most of the regions studied, there is a substantial number of 
very small fragments. This is not consistent with the fact that the 
smallest fragments should be photo-evaporated the fastest and points to an active 
fragment production process. Fragment size distributions,
such as the ones in Fig.~\ref{fig:sizes_all}, can provide constraints to theories of 
how molecular cloud fragments are produced.

\item Tear-shaped cusps and circular silhouettes can have similar
sizes (where chords are compared to diameters). We therefore suspect that
these molecular cloud fragments belong to the same class, where the former
is fully exposed to the ionizing radiation field, while the latter is shielded from it.
This is, however, a ball-park estimate argument and only a more accurate model could 
properly address this point.

\item From the analysis of the optical depth properties of the 
circular silhouettes of \citet{Rei+03}, we determined that the circular
silhouettes in IC~2944 and possibly those in M~16 
cannot be self-gravitating and collapsing to make stars. Using their
assumptions about the optical depth of the fragments, we calculate 
Jeans 
lengths 
which are much larger than the fragment themselves, therefore pointing to their transient nature.
We argue, however, that
the {\it assumption} might actually be flawed and that the fragments might actually
be rather centrally concentrated and possibly star-forming.
Only in this way, would we explain the circular shapes of the silhouette fragments.

\item Only one silhouette (in M~17) out of {\it all} the features found in the eight regions
might have a central star. If this feature is indeed a proplyd (some reservations exist),
it joins a small number of {\it bona fide} proplyds
found outside the Orion Nebula. Since several of the observed regions are close enough to enable us 
to resolve the size of the largest proplyds found in the Orion Nebula, one might wonder why
so few proplyds are found. Recent work by Smith et al. (2005) might provide an explanation.
According to them,
proplyds are observed only when they are close to bright illuminating sources.
However this proximity tends to evaporate them on short time scales. As a result we only see
the proplyds when low mass stars have formed near massive stars {\it and} we observe them
at the very beginning of their lives (within the first 500\,000 years). The cusps we find near
the cavity walls, might have been recently exposed by the advancement of the ionization front.
Their placental material is externally ionized and evaporated giving them a limb-brightened
appearance. Eventually, the nascent star in their middle will be completely stripped 
(surrounded by a disk or not) but will 
not shine as a proplyd because it is not sufficiently close to the ionizing source
(for further elaborations on this point, the reader is refereed to \citealt{Joh+98}).

\end{itemize}

\acknowledgments

We are thankful to David Zurek for spotting fragment \#12 in M~17, a
possible {\it bona fide} circumstellar disk around a young star and for 
carrying out the early data reduction on the M~17 ERO observations that 
helped us write the proposal that lead to the current dataset.
We acknowledge useful discussion with Hector Harce and Mordecai-Mark Mac Low.
We thank Katie Ray and Geoff Clayton for reading and commenting on the manuscript
and the Referee, Will Henney, for comments that improved the paper.
RHR acknowledges support from the NASA Long-Term Space Astrophysics 
(LTSA) program. This work was supported by an 
HST grant associated with
proposal GO-9857.
This research has made use of the SIMBAD database,
operated at CDS, Strasbourg, France


\begin{thebibliography}{50}
\expandafter\ifx\csname natexlab\endcsname\relax\def\natexlab#1{#1}\fi


\bibitem[{{Abt} \& {Corbally}(2000)}]{AC00}
{Abt}, H.~A. \& {Corbally}, C.~J. 2000, \apj, 541, 841
\bibitem[{{Allen}(1986)}]{All86}
{Allen}, D.~A. 1986, \mnras, 219, 35P

\bibitem[{{Ardeberg} \& {Maurice}(1980)}]{AM80}
{Ardeberg}, A. \& {Maurice}, E. 1980, \aaps, 39, 325

\bibitem[{{Bally} {et~al.}(2005){Bally}, {Licht}, {Smith}, \&
  {Walawender}}]{Bal+05}
{Bally}, J., {Licht}, D., {Smith}, N., \& {Walawender}, J. 2005, \aj, 129, 355

\bibitem[{{Bally} {et~al.}(2000){Bally}, {O'Dell}, \& {McCaughrean}}]{Bal+00}
{Bally}, J., {O'Dell}, C.~R., \& {McCaughrean}, M.~J. 2000, \aj, 119, 2919

\bibitem[{{Bally} {et~al.}(1998){Bally}, {Yu}, {Rayner}, \&
  {Zinnecker}}]{Bal+98}
{Bally}, J., {Yu}, K.~C., {Rayner}, J., \& {Zinnecker}, H. 1998, \aj, 116, 1868

\bibitem[{{Belikov} {et~al.}(1999){Belikov}, {Kharchenko}, {Piskunov}, \&
  {Schilbach}}]{Bel+99}
{Belikov}, A.~N., {Kharchenko}, N.~V., {Piskunov}, A.~E., \& {Schilbach}, E.
  1999, \aaps, 134, 525

\bibitem[{{Belikov} {et~al.}(2000){Belikov}, {Kharchenko}, {Piskunov}, \&
  {Schilbach}}]{Bel+00}
---. 2000, \aap, 358, 886

\bibitem[{{Bica} {et~al.}(2003){Bica}, {Dutra}, \& {Barbuy}}]{Bic+03}
{Bica}, E., {Dutra}, C.~M., \& {Barbuy}, B. 2003, \aap, 397, 177

\bibitem[{{Bok} \& {Reilly}(1947)}]{BR47}
{Bok}, B.~J. \& {Reilly}, E.~F. 1947, \apj, 105, 255

\bibitem[{{Brandner} {et~al.}(2000){Brandner}, {Grebel}, {Chu}, {Dottori},
  {Brandl}, {Richling}, {Yorke}, {Points}, \& {Zinnecker}}]{Bra+00}
{Brandner}, W., {Grebel}, E.~K., {Chu}, Y.-H., {Dottori}, H., {Brandl}, B.,
  {Richling}, S., {Yorke}, H.~W., {Points}, S.~D., \& {Zinnecker}, H. 2000,
  \aj, 119, 292

\bibitem[{{Carraro} {et~al.}(2001){Carraro}, {Patat}, \& {Baumgardt}}]{Car+01}
{Carraro}, G., {Patat}, F., \& {Baumgardt}, H. 2001, \aap, 371, 107

\bibitem[{{Clemens} {et~al.}(1991){Clemens}, {Yun}, \& {Heyer}}]{CYH91}
{Clemens}, D.~P., {Yun}, J.~L., \& {Heyer}, M.~H. 1991, \apjs, 75, 877

\bibitem[{{Cruz-Gonzalez} {et~al.}(1974){Cruz-Gonzalez}, {Recillas-Cruz},
  {Costero}, {Peimbert}, \& {Torres-Peimbeert}}]{Cru+74}
{Cruz-Gonzalez}, C., {Recillas-Cruz}, E., {Costero}, R., {Peimbert}, M., \&
  {Torres-Peimbeert}, S. 1974, Revista Mexicana de Astronomia y Astrofisica, 1,
  211

\bibitem[{{Dutra} {et~al.}(2003){Dutra}, {Bica}, {Soares}, \&
  {Barbuy}}]{Dut+03}
{Dutra}, C.~M., {Bica}, E., {Soares}, J., \& {Barbuy}, B. 2003, \aap, 400, 533

\bibitem[{{Feinstein} \& {Vazquez}(1989)}]{FV89}
{Feinstein}, A. \& {Vazquez}, R.~A. 1989, \aaps, 77, 321

\bibitem[{{Guetter} \& {Turner}(1997)}]{GT97}
{Guetter}, H.~H. \& {Turner}, D.~G. 1997, \aj, 113, 2116

\bibitem[{{Hanson} {et~al.}(1997){Hanson}, {Howarth}, \& {Conti}}]{HHC97}
{Hanson}, M.~M., {Howarth}, I.~D., \& {Conti}, P.~S. 1997, \apj, 489, 698

\bibitem[{{Henney} \& {O'Dell}(1999)}]{HO99}
{Henney}, W.~J. \& {O'Dell}, C.~R. 1999, \aj, 118, 2350

\bibitem[{{Henney} {et~al.}(2002){Henney}, {O'Dell}, {Meaburn}, {Garrington},
  \& {Lopez}}]{Hen+02}
{Henney}, W.~J., {O'Dell}, C.~R., {Meaburn}, J., {Garrington}, S.~T., \&
  {Lopez}, J.~A. 2002, \apj, 566, 315

\bibitem[{{Hester} {et~al.}(1996){Hester}, {Scowen}, {Sankrit}, {Lauer},
  {Ajhar}, {Baum}, {Code}, {Currie}, {Danielson}, {Ewald}, {Faber},
  {Grillmair}, {Groth}, {Holtzman}, {Hunter}, {Kristian}, {Light}, {Lynds},
  {Monet}, {O'Neil}, {Shaya}, {Seidelmann}, \& {Westphal}}]{Hes+96}
{Hester}, J.~J., {Scowen}, P.~A., {Sankrit}, R., {Lauer}, T.~R., {Ajhar},
  E.~A., {Baum}, W.~A., {Code}, A., {Currie}, D.~G., {Danielson}, G.~E.,
  {Ewald}, S.~P., {Faber}, S.~M., {Grillmair}, C.~J., {Groth}, E.~J.,
  {Holtzman}, J.~A., {Hunter}, D.~A., {Kristian}, J., {Light}, R.~M., {Lynds},
  C.~R., {Monet}, D.~G., {O'Neil}, E.~J., {Shaya}, E.~J., {Seidelmann}, K.~P.,
  \& {Westphal}, J.~A. 1996, \aj, 111, 2349

\bibitem[{{Jeans}(1902)}]{Jea1902}
{Jeans}, J.~H. 1902, Philos.\ Trans.\ R.\ Soc.\ London A, 199, 1

\bibitem[{{Johnstone} {et~al.}(1998){Johnstone}, {Hollenbach}, \&
  {Bally}}]{Joh+98}
{Johnstone}, D., {Hollenbach}, D., \& {Bally}, J. 1998, \apj, 499, 758

\bibitem[{{Joye} \& {Mandel}(2003)}]{JM03}
{Joye}, W.~A. \& {Mandel}, E. 2003, in ASP Conf. Ser. 295: Astronomical Data
  Analysis Software and Systems XII, 489--+

\bibitem[{{Lod{\'e}n}(1966)}]{Lod66}
{Lod{\'e}n}, L.~O. 1966, Arkiv for Astronomi, 4, 65

\bibitem[{{Ma{\'{\i}}z-Apell{\'a}niz}
  {et~al.}(2004){Ma{\'{\i}}z-Apell{\'a}niz}, {Walborn}, {Galu{\'e}}, \&
  {Wei}}]{Mai+04}
{Ma{\'{\i}}z-Apell{\'a}niz}, J., {Walborn}, N.~R., {Galu{\'e}}, H.~{\'A}., \&
  {Wei}, L.~H. 2004, \apjs, 151, 103

\bibitem[{{Massey} {et~al.}(2001){Massey}, {DeGioia-Eastwood}, \&
  {Waterhouse}}]{Mas+01}
{Massey}, P., {DeGioia-Eastwood}, K., \& {Waterhouse}, E. 2001, \aj, 121, 1050

\bibitem[{{McCaughrean} \& {O'Dell}(1996)}]{MO96}
{McCaughrean}, M.~J. \& {O'Dell}, C.~R. 1996, \aj, 111, 1977

\bibitem[{{O'Dell}(2001)}]{ODe01}
{O'Dell}, C.~R. 2001, \araa, 39, 99

\bibitem[{{O'Dell} {et~al.}(1993){O'Dell}, {Wen}, \& {Hu}}]{ODe+93}
{O'Dell}, C.~R., {Wen}, Z., \& {Hu}, X. 1993, \apj, 410, 696

\bibitem[{{Paladini} {et~al.}(2003){Paladini}, {Burigana}, {Davies}, {Maino},
  {Bersanelli}, {Cappellini}, {Platania}, \& {Smoot}}]{Pal+03}
{Paladini}, R., {Burigana}, C., {Davies}, R.~D., {Maino}, D., {Bersanelli}, M.,
  {Cappellini}, B., {Platania}, P., \& {Smoot}, G. 2003, \aap, 397, 213
\bibitem[{{Pellerin} {et~al.}(2002){Pellerin}, {Fullerton}, {Robert}, {Howk},
  {Hutchings}, {Walborn}, {Bianchi}, {Crowther}, \& {Sonneborn}}]{Pel+02}
{Pellerin}, A., {Fullerton}, A.~W., {Robert}, C., {Howk}, J.~C., {Hutchings},
  J.~B., {Walborn}, N.~R., {Bianchi}, L., {Crowther}, P.~A., \& {Sonneborn}, G.
  2002, \apjs, 143, 159

\bibitem[{{Reid} {et~al.}(1991){Reid}, {Brewer}, {Brucato}, {McKinley},
  {Maury}, {Mendenhall}, {Mould}, {Mueller}, {Neugebauer}, {Phinney},
  {Sargent}, {Schombert}, \& {Thicksten}}]{Rei+91}
{Reid}, I.~N., {Brewer}, C., {Brucato}, R.~J., {McKinley}, W.~R., {Maury}, A.,
  {Mendenhall}, D., {Mould}, J.~R., {Mueller}, J., {Neugebauer}, G., {Phinney},
  J., {Sargent}, W.~L.~W., {Schombert}, J., \& {Thicksten}, R. 1991, \pasp,
  103, 661

\bibitem[{{Reipurth} {et~al.}(1997){Reipurth}, {Corporon}, {Olberg}, \&
  {Tenorio-Tagle}}]{Rei+97}
{Reipurth}, B., {Corporon}, P., {Olberg}, M., \& {Tenorio-Tagle}, G. 1997,
  \aap, 327, 1185

\bibitem[{{Reipurth} {et~al.}(2003){Reipurth}, {Raga}, \& {Heathcote}}]{Rei+03}
{Reipurth}, B., {Raga}, A., \& {Heathcote}, S. 2003, \aj, 126, 1925
\bibitem[{{Schild} {et~al.}(1983){Schild}, {Garrison}, \& {Hiltner}}]{Sch+83}
{Schild}, R.~E., {Garrison}, R.~F., \& {Hiltner}, W.~A. 1983, \apjs, 51, 321

\bibitem[{{Smith} {et~al.}(2003){Smith}, {Bally}, \& {Morse}}]{Smi+03}
{Smith}, N., {Bally}, J., \& {Morse}, J.~A. 2003, \apjl, 587, L105

\bibitem[{{Smith} {et~al.}(2005){Smith}, {Bally}, {Shuping}, {Morris}, \&
  {Kassis}}]{Smi+05b}
{Smith}, N., {Bally}, J., {Shuping}, R.~Y., {Morris}, M., \& {Kassis}, M. 2005,
  \aj, 130, 1763

\bibitem[{{Stapelfeldt} {et~al.}(1997){Stapelfeldt}, {Sahai}, {Werner}, \&
  {Trauger}}]{Sta+97}
{Stapelfeldt}, K., {Sahai}, R., {Werner}, M., \& {Trauger}, J. 1997, in ASP
  Conf. Ser. 119: Planets Beyond the Solar System and the Next Generation of
  Space Missions, 131--+

\bibitem[{{Stecklum} {et~al.}(1998){Stecklum}, {Henning}, {Feldt}, {Hayward},
  {Hoare}, {Hofner}, \& {Richter}}]{Ste+98}
{Stecklum}, B., {Henning}, T., {Feldt}, M., {Hayward}, T.~L., {Hoare}, M.~G.,
  {Hofner}, P., \& {Richter}, S. 1998, \aj, 115, 767
\bibitem[{{Sung} {et~al.}(2000){Sung}, {Chun}, \& {Bessell}}]{Sun+00}
{Sung}, H., {Chun}, M.-Y., \& {Bessell}, M.~S. 2000, \aj, 120, 333

\bibitem[{{Thackeray}(1950)}]{Tha50}
{Thackeray}, A.~D. 1950, \mnras, 110, 524

\bibitem[{{Tody}(1986)}]{Tod86}
{Tody}, D. 1986, in Instrumentation in astronomy VI; Proceedings of the
  Meeting, Tucson, AZ, Mar. 4-8, 1986. Part 2 (A87-36376 15-35). Bellingham,
  WA, Society of Photo-Optical Instrumentation Engineers, 1986, p. 733., 733--+

\bibitem[{{Tody}(1993)}]{Tod93}
{Tody}, D. 1993, in ASP Conf. Ser. 52: Astronomical Data Analysis Software and
  Systems II, 173--+

\bibitem[{{van den Ancker} {et~al.}(1997){van den Ancker}, {The}, {Feinstein},
  {Vazquez}, {de Winter}, \& {Perez}}]{Van+97}
{van den Ancker}, M.~E., {The}, P.~S., {Feinstein}, A., {Vazquez}, R.~A., {de
  Winter}, D., \& {Perez}, M.~R. 1997, \aaps, 123, 63

\bibitem[{{Vannier} {et~al.}(2001){Vannier}, {Lemaire}, {Field}, {Pineau des
  For{\^ e}ts}, {Pijpers}, \& {Rouan}}]{Van+01}
{Vannier}, L., {Lemaire}, J.~L., {Field}, D., {Pineau des For{\^ e}ts}, G.,
  {Pijpers}, F.~P., \& {Rouan}, D. 2001, \aap, 366, 651

\bibitem[{{V{\'a}zquez-Semadeni} {et~al.}(2005){V{\'a}zquez-Semadeni}, {Kim},
  {Shadmehri}, \& {Ballesteros-Paredes}}]{Vaz+05}
{V{\'a}zquez-Semadeni}, E., {Kim}, J., {Shadmehri}, M., \&
  {Ballesteros-Paredes}, J. 2005, \apj, 618, 344

\bibitem[{{Walborn}(1972)}]{Wal72}
{Walborn}, N.~R. 1972, \aj, 77, 312

\bibitem[{{Walborn}(1982)}]{Wal82}
---. 1982, \aj, 87, 1300

\bibitem[{{Walborn}(1987)}]{Wal87}
---. 1987, \aj, 93, 868

\bibitem[{{Warren} \& {Hesser}(1977)}]{WH77}
{Warren}, W.~H. \& {Hesser}, J.~E. 1977, \apjs, 34, 115

\bibitem[{{Yusef-Zadeh} {et~al.}(2005){Yusef-Zadeh}, {Biretta}, \&
  {Geballe}}]{Yus+05}
{Yusef-Zadeh}, F., {Biretta}, J., \& {Geballe}, T.~R. 2005, \aj, 130, 1171

\end{thebibliography}

\clearpage

\clearpage


\clearpage


\clearpage


\clearpage


\clearpage


\clearpage

\clearpage 

\clearpage

\clearpage

\clearpage 

\clearpage

\begin{deluxetable}{ccccccc}
\tabletypesize{\scriptsize}
\tablecaption{Observing log.\label{tab:observations}}
\tablehead{
\colhead{Object} & \colhead{RA \& Dec} & PA & \colhead{Instrument} & \colhead{Filter} & \colhead{UT Date} & \colhead{Exposure time} \\
\colhead{}       & \colhead{(J2000)} & (deg) & \colhead{}           & \colhead{}       &  \colhead{}        & \colhead{(sec)}  
}
\startdata
\multicolumn{7}{c}{Complete image set from program GO~9857}\\
NGC~6530-V1 & 18 04 54.87 -24 17 17.3 &299.9& WFPC2   & F656N  & 2004-04-04 & 525\\
NGC~6530-V1 & 18 04 31.50 -24 21 15.7 &255.0& ACS/WFC & F658N  & 2004-04-04 & 780\\
NGC~6530-V1 & 18 04 31.50 -24 21 15.7 &255.0& ACS/WFC & F660N  & 2004-04-04 & 800\\
NGC~6530-V1 & 18 04 31.50 -24 21 15.7 &255.0& ACS/WFC & F550N  & 2004-04-04 & 400\\
NGC~6530-V2 & 18 04 13.21 -24 21 40.0 &104.8& WFPC2   & F656N  & 2003-07-07 & 525   \\
NGC~6530-V2 & 18 03 49.20 -24 19 53.0 &60.0 & ACS/WFC & F658N  & 2003-07-07 & 780\\
NGC~6530-V2 & 18 03 49.20 -24 19 53.0 &60.0 & ACS/WFC & F660N  & 2003-07-07 & 800   \\
NGC~6530-V2 & 18 03 49.20 -24 19 53.0 &60.0 & ACS/WFC & F550N  & 2003-07-07 & 400   \\
NGC~3324    & 10 36 51.40 -58 40 43.9 &214.6& WFPC2   & F656N  & 2004-08-19 & 851   \\
NGC~3324    & 10 37 18.50 -58 36 11.0 &169.8& ACS/WFC & F658N  & 2004-08-19 & 1300  \\
NGC~3324    & 10 37 18.50 -58 36 11.0 &169.8& ACS/WFC & F660N  & 2004-08-19 & 1960  \\
NGC~3324    & 10 37 18.50 -58 36 11.0 &169.8& ACS/WFC & F550N  & 2004-08-19 & 654   \\
NGC~2467    & 07 52 44.35 -26 25 12.5 &79.8 & WFPC2   & F656N  & 2004-02-18 & 410 \\
NGC~2467    & 07 52 18.20 -26 26 06.0 &35.0 & ACS/WFC & F658N  & 2004-02-18 & 750 \\
NGC~2467    & 07 52 18.20 -26 26 06.0 &35.0 & ACS/WFC & F660N  & 2004-02-18 & 897 \\
NGC~2467    & 07 52 18.20 -26 26 06.0 &35.0 & ACS/WFC & F550N  & 2004-02-18 & 340 \\
Pismis~24   & 17 25 09.30 -34 12 15.7 &114.4& WFPC2   & F656N  & 2003-07-08 & 505    \\
Pismis~24   & 17 24 43.60 -34 09 40.5 &69.6 & ACS/WFC & F658N  & 2003-07-08 & 685 \\
Pismis~24   & 17 24 43.60 -34 09 40.5 &69.6 & ACS/WFC & F660N  & 2003-07-08 & 960 \\
Pismis~24   & 17 24 43.60 -34 09 40.5 &69.6 & ACS/WFC & F550N  & 2003-07-08 & 350 \\
IC~1590     & 00 53 30.64  56 37 30.1 &76.0 & WFPC2   & F656N  & 2004-03-02 & 605    \\
IC~1590     & 00 52 49.00  56 36 12.7 &31.1 & ACS/WFC & F658N  & 2004-03-02 & 678    \\
IC~1590     & 00 52 49.00  56 36 12.7 &31.1 & ACS/WFC & F660N  & 2004-03-02 & 1017   \\
IC~1590     & 00 52 49.00  56 36 12.7 &31.1 & ACS/WFC & F550M  & 2004-03-02 & 450    \\
\multicolumn{7}{c}{Selected archival images used}\\
IC~2944     & 11 38 22.25 -63 20 37.2 &14.8 & WFPC2   & F656N  & 1999-02-07 & 452\tablenotemark{a}       \\
M~16    & 18 18 42.13 -13 47 58.6 &143.8& WFPC2   & F656N  & 2002-08-08 & 2600\tablenotemark{b}      \\
M~17        & 18 20 34.91 -16 09 27.54&267.1& ACS/WFC & F658N  & 2002-04-01 & 760       \\
M~17        & 18 20 34.91 -16 09 27.54&267.1& ACS/WFC & F502N  & 2002-04-02 & 760       \\
M~17        & 18 20 34.91 -16 09 27.54&267.1& ACS/WFC & F435W  & 2002-04-01 & 710       \\
M~17        & 18 20 34.91 -16 09 27.54&267.1& ACS/WFC & F814W  & 2002-04-01 & 720       \\
\enddata
\tablenotetext{a}{From averaging the eight available H$\alpha$ images.}
\tablenotetext{b}{From averaging the four available H$\alpha$ images.}
\end{deluxetable}

\begin{deluxetable}{lllllllll}
\tabletypesize{\scriptsize}
\tablecaption{Cluster/\ion{H}{2} region parameters.\label{tab:clusters}}
\tablehead{
\colhead{Name} & \colhead{RA \& Dec\tablenotemark{a}} & \colhead{Distance} & \colhead{Age}   & 
\colhead{Diameter} & \colhead{O Star\tablenotemark{b}}  & \colhead{Earliest} 
& \colhead{FOV} & \colhead{3-pixels\tablenotemark{c}} \\
\colhead{} & \colhead{} & \colhead{$\pm$error} & \colhead{[age spread]}   & 
\colhead{HII} & \colhead{content}  & \colhead{spectral} &&\\
\colhead{}     & \colhead{(J2000)}     &  \colhead{(kpc)}   & \colhead{(Myr)} & 
\colhead{(pc)}&  \colhead{}   & \colhead{type} & \colhead{(pc)} & \colhead{(AU)}            
}
\startdata
NGC~6530\tablenotemark{d}   &18 04 48 -24 20   &  1.8$\pm$0.1            &1.5[4.0]    & 18   & 3E/2L & O4~V & 1.8 & 270 \\
NGC~3324\tablenotemark{e}   &10 37 18 -58 38   &  3.0$\pm$0.1            &$<$2-3      & 12   & 1E/1L & O6.5~V & 3.0 & 450 \\
NGC~2467                    &07 52 19 -26 26 30&  4.1\tablenotemark{f}   & 12::\tablenotemark{g}&8&1E/1L&O6~V & 4.1 & 610\\
Pismis~24\tablenotemark{h} &17 25 24 -34 26   &  2.5   &0.7[2.3]    &50/2.5& 4E/4L & O3~I,III & 2.5 & 370\\
IC~1590\tablenotemark{j}    &00 53 03 +56 35   &  2.94$\pm$0.15          &3.5$\pm$0.2 &21    & 1E/2L & O6.5~V & 2.9 & 440\\
IC~2944\tablenotemark{k}    &11 38 20 -63 22 22&  1.8\tablenotemark{l}   &7           &28    &4E/5L  & O6~V & 0.60& 230\\
M~16                    &18 18 48 -13 47   &  2.14$\pm$0.10\tablenotemark{m}   &6[6]\tablenotemark{n}&18&3E/6L&O4~V&0.87&330\\
M~17\tablenotemark{o}       &18 20 26 -16 10 36&  1.3                    &1           &37    &2E/2L  &O5  & 1.3 &190\\
\enddata

\tablenotetext{a}{These are the ICRS coordinates from the SIMBAD database.}
\tablenotetext{b}{``E" stands for Early (O3-O6), ``L" stands for late (O7-O9). 
The number indicates how many early or late O stars are present in the region.}
\tablenotetext{c}{Size of 3-pixels in AU. This is the minimum detection size of a fragment.}
\tablenotetext{d}{Sung et al. 2000.}
\tablenotetext{e}{Carraro et al. 2001.}
\tablenotetext{f}{Feinstein \& Vasquez 1989.}
\tablenotetext{g}{\citealp{Lod66}; very uncertain.}
\tablenotetext{h}{Massey et al. 2001.}
\tablenotetext{j}{\citealp{GT97}.}
\tablenotetext{k}{\citealp{Wal87}.}
\tablenotetext{l}{\citealp{AM80}.}
\tablenotetext{m}{\citealp{Bel+99}.}
\tablenotetext{n}{\citealp{Bel+00}.}
\tablenotetext{o}{\citealp{HHC97}.}
\end{deluxetable}

\begin{deluxetable}{lllll}
\tabletypesize{\scriptsize}
\tablewidth{0pt}
\tablecaption{O stars in the observed regions.\label{tab:stars}}
\tablehead{
\colhead{Name (Alias)}  & \colhead{RA \& Dec}  &  \colhead{V Magnitudes} & \colhead{Spectral}  &  \colhead{Reference} \\
                        & \colhead{(J2000)}  & \colhead{(mag)}   & \colhead{Type\tablenotemark{a}}      &        \\
}
\startdata
\multicolumn{5}{c}{NGC~6530} \\
 HD164740 (Her 36)    & 18 03 40.2 -24 22 43 & 9.1   & O7.5~V      & \citealt{Wal82} \\
 HD164794 (9 Sgr)     & 18 03 52.4 -24 21 38 & 5.96   & O4~V((f))  & \citealt{Wal72,Sch+83} \\
 HD164816             & 18 03 56.9 -24 18 45 & 7.09  & O9.5~III-IVn& \citealt{Mai+04} \\
 HD166052             & 18 05 10.5 -24 23 54 & 6.9   & O6.5+O6.5~V & \citealt{Van+97} \\
\multicolumn{5}{c}{NGC~3324} \\
 HD092206AB           & 10 37 22.3 -58 37 23 & 7.82  & O6.5~V(n)    & \citealt{Mai+04}\\
 HD092206C            & 10 37 18.6 -58 37 42 & -     & O8.5~Vp      & \citealt{Mai+04}\\
\multicolumn{5}{c}{NGC~2467} \\
 HD64315              & 07 52 20.3 -26 25 46  & 9.19      & O6~Vn         & Walborn 1982\\
 CD-26 5129           & 07 52 55.4 -26 28 42  & 9.39      & O7           & Feinstein \& Vazquez 1989\\
\multicolumn{5}{c}{Pismis~24} \\
 Cl Pismis 24 15      & 17 24 28.7 -34 14 50 &  12.32   & O8~V       &Massey et al. 2001\\    
 Cl Pismis 24 10      & 17 24 35.9 -34 13 59 &  13.02   & O9~V       &Massey et al. 2001\\    
 Cl Pismis 24 3       & 17 24 42.2 -34 13 21 &  12.75   & O8~V       &Massey et al. 2001\\    
 HD319718             & 17 24 42.9 -34 11 48  &  10.43   & O3~If*         &Massey et al. 2001\\
 Cl Pismis 24 2       & 17 24 43.2 -34 12 43 &  11.95   & O5.5~V((f))     &Massey et al. 2001\\    
 Cl Pismis 24 16 (N58)& 17 24 44.3 -34 12 00  &  13.02   & O7.5~V       &Massey et al. 2001\\
 Cl Pismis 24 17 (N57)& 17 24 44.7 -34 12 02  &  11.84   & O3~III(f*)   &Massey et al. 2001\\    
 Cl Pismis 24 13 (N36)& 17 24 45.7 -34 09 39 &  12.73   & O6.5~V((f)) &Massey et al. 2001\\    
 HD157504 (WR 93)     & 17 25 08.9 -34 11 12 &  11.46   & WC7+abs?    & Massey et al. 2001\\
\multicolumn{5}{c}{IC~1590\tablenotemark{b}} \\
 HD5005AB            & 00 52 49.2 +56 37 39   & 8.34     & O6.5~V((f)) & \citealt{GT97}\\
 HD5005C             & 00 52 49.4 +56 37 39   & 9.19     & O8~Vn & Guetter \& Turner 1997\\
 HD5005D             & 00 52 49.6 +56 37 36   & 9.78     & O9~Vn & Guetter \& Turner 1997\\
\multicolumn{5}{c}{IC~2944} \\
 HD101131 & 11 37 48.4 -63 19 23 & 7.14 & O6~V((f))    & \citealt{Mai+04}\\
 HD308813 & 11 37 58.4 -63 18 59 & 9.28 & O9.5~V       & \citealt{Pel+02}\\
 HD101190 & 11 38 09.9 -63 11 48 & 7.31 & O6~V((f))    & \citealt{Mai+04}\\
 HD101191 & 11 38 12.2 -63 23 26 & 8.49 & O8~V((n))    & \citealt{Mai+04}\\
 HD101205 & 11 38 20.4 -63 22 21 & 6.46 & O7~IIIn((f)) & \citealt{Mai+04}\\
 HD101223 & 11 38 22.8 -63 12 02 & 8.69 & O8~V((f))    & \citealt{Mai+04}\\
 HD101298 & 11 39 03.3 -63 25 47 & 8.07 & O6~V((f))    & \citealt{Mai+04}\\
 HD101413 & 11 39 45.8 -63 28 40 & 8.35 & O8~V         & \citealt{Mai+04}\\
 HD101436 & 11 39 49.9 -63 28 43 & 7.59 & O6.5~V       & \citealt{Mai+04}\\
\multicolumn{5}{c}{M~16} \\
 BD-13 4921            & 18 18 29.9 -13 49 57 & 9.85 &O9~V    & \citealt{Cru+74}\\
 NGC 6611 166          & 18 18 32.2 -13 48 48 & 10.35 & O9~V & \citealt{Cru+74}\\
 BD-13 4923            & 18 18 32.7 -13 45 12 & 10.08& O6  &  \citealt{Cru+74}\\ 
 HD168075              & 18 18 36.1 -13 47 36 & 8.75 & O6~V((f)) & \citealt{Mai+04}\\
 HD168076              & 18 18 36.4 -13 48 02 & 8.20 & O4~V((f)) & \citealt{Mai+04}\\
 BD-13 4927            & 18 18 40.1 -13 45 18 & 9.55 & O7~Ib(f)  & \citealt{Mai+04}\\
 BD-13 4929            & 18 18 45.9 -13 46 31 &  9.86 & O9.5~V &  \citealt{Cru+74}\\ 
 BD-13 4930            & 18 18 52.7 -13 49 42 & 9.44 & O9.5~V &  \citealt{Cru+74}\\ 
 HD168137              & 18 18 56.2 -13 48 31 & 8.95 & O8~V   &  \citealt{Cru+74}\\ 
\multicolumn{5}{c}{M~17} \\
 Cl* NGC 6618 B 260 & 18 20 26.0 -16 08 32 & 14.20 & O7-8~V & Hanson et al. 1997\\
 Cl* NGC 6618 B 189 & 18 20 29.8 -16 10 44 & 14.13 & O5~V   & Hanson et al. 1997\\
 Cl* NGC 6618 B 111 & 18 20 34.5 -16 10 12 & 11.29 & O5~V   & Hanson et al. 1997\\
 Cl* NGC 6618 B 98  & 18 20 35.4 -16 10 48 & 10.1  & O9~V   & Hanson et al. 1997\\
\enddata
\tablenotetext{a}{The spectral type notation, originally from \citet{Wal72}, is also explained in the listed references.}
\tablenotetext{b}{There is a great deal of confusion about how many stars reside in this tight cluster. For instance,
one would believe that Tr~13, cited by \citet{AC00} is an additional star but this star, also called ADS719, appears
cross-referenced to HD5005 in the SIMBAD database. Therefore we only list the stars from \citet{GT97}, cautioning that
this might provide only a lower limit to the stellar content and number of hard photons in this region.}
\end{deluxetable}

\begin{deluxetable}{lllllll}
\tabletypesize{\scriptsize}
\tablewidth{0pt}
\tablecaption{Discrete features detected in the observed regions.\label{tab:features}}
\tablehead{
\colhead{ID} & \colhead{Class}  & 
\colhead{Shape} &  \colhead{Edges} & 
\colhead{Location}  &  \colhead{Short \& Long\tablenotemark{a}}
& \colhead{(X,Y)\tablenotemark{b}}\\
\colhead{} & \colhead{}  & \colhead{}  &  \colhead{} & \colhead{}  &  \colhead{dimensions} & \colhead{position}\\
             &                  &                  &                 &                     &  \colhead{(AU)}        & \colhead{(pix)}
}
\startdata
\multicolumn{7}{l}{NGC~6530} \\                                                                    
1    & cusp     & -             & LB       & attached ridge  &     1530      2700 &     2808      1336 \\
2    & cusp     & -             & LB       & attached ridge  &      990      3600 &     2761      1313 \\
3    & cusp     & -             & LB       & attached ridge  &      900      3330 &     2764       731 \\
4    & cusp     & rounded       & LB       & attached ridge  &     2700      9001 &     2901      1038 \\
  \multicolumn{7}{l}{NGC~2467} \\                                                                    
1    &fragment  & complex       & mix      & isolated        &     6561     38954 &      333      2349 \\
2    & sil.     & irregular     & fuzzy    & near frag. 4    &     2050     21732 &      741      3536 \\
3    & sil.     & nodule        & sharp    & near frag. 4    &     1025           &      613      3918 \\
4    &fragment  & complex       & mix      & near frag. 7    &    33828     59045 &      864      4006 \\
5    & cusps    & rounded       & LB       & attached frag. 4&     1763      2870 &      865      3923 \\
6    & cusp     & rounded       & LB       & attached frag. 4&     4510      8611 &     1009      3956 \\
7    &fragment  & complex       & mix      & near frag. 4    &    20502     45104 &     1300      3910 \\
8    & sil.     & -             & fuzzy    & attached frag. 7&     2952      5330 &     1390      3743 \\
9    & cusp     & rounded       & LB       & near frag. 7    &      820           &     1646      3679 \\
10   &fragment  & complex       & mix      & near frag. 11   &    11071     29113 &     1847      3445 \\
11   &fragment  & complex       & mix      & near frag. 16   &    20502     94309 &     1857      3225 \\
12   & cusp     & rounded       & LB       & attached frag.11&     1435      9841 &     1866      3128 \\
13   & cusp     & rounded       & LB       & near frag. 11   &     4100           &     2032      3301 \\
14   &fragment  & complex       & mix      & near frag. 16   &     8201     22142 &     2325      3336 \\
15   & cusp     & rounded       & LB       & attached frag.16&     4100      7381 &     2515      3344 \\
16   &fragment  & complex       & mix      & near frag. 26   &     6151     61506 &     2603      3259 \\
17   & sil.     & rounded       & fuzzy    & isolated        &     3403           &     2440      2219 \\
18   & sil.     & rounded       & sharp    & isolated        &     2132           &     2493      2314 \\
19   & sil.     & triangular    & sharp    & near frag. 22   &     6561           &     2535      2465 \\
20   & sil.     & nodule        & sharp    & near frag. 22   &     1804           &     2693      2486 \\
21   & sil.     & complex       & sharp    & attached frag.22&     4510     32803 &     2627      2631 \\
22   &fragment  & complex       & mix      & isolated        &    10251     73807 &     2779      2773 \\
23   & cusp     & complex       & LB       & isolated        &     5741           &     3061      2483 \\
24   & cusp     & -             & LB       & attached ridge  &     2132      8201 &     3491      2275 \\
25   & cusp     & complex       & LB       & near frag. 26   &     4920           &     2757      3409 \\
26   &fragment  & complex       & mix      & attached ridge  &    40184    102509 &     3215      3573 \\
27   &fragment  & complex       & LB       & attached frag.26&     4920     17222 &     2956      3660 \\
28   & cusp     & -             & LB       & attached frag.26&     4510     13531 &     3303      3435 \\
29   & cusp     & hooked        & LB       & attached frag.26&     2050      5741 &     3375      3652 \\
30   & cusp     & rounded       & LB       & attached ?      &     5741      7791 &     3901      3774 \\
31   &fragment  & complex       & mix      & near ridge      &    28703     41004 &     3930      3602 \\
32   & cusp     & -             & LB       & near ridge      &     5330     25422 &     3981      2815 \\
33   & cusp     & -             & LB       & near ridge      &     1845           &     3901      2615 \\
34   & sil.     & nodule        & sharp    & near ridge      &     1148           &      459       596 \\
35   & cusp     & -             & LB       & attached ridge  &     1230      6151 &      319      1064 \\
36   & cusp     & -             & LB       & attached ridge  &     2870     28293 &      460      1154 \\
37   & cusp     & -             & LB       & attached ridge  &     6971     15171 &      606      1041 \\
38   & cusp     & -             & LB       & attached ridge  &     4510     22552 &      732      1173 \\
39   & cusp     & multi-tip     & sharp    & near ridge      &     5741     14761 &      605      1328 \\
40   &fragment  & complex       & mix      & attached ridge  &    26652     82007 &      436      1408 \\
41   & cusp     & w/ jet        & LB       & attached ridge  &     1271      5741 &      582      1474 \\
42   & cusp     & -             & LB       & attached frag.  &     4920     10661 &      571      1599 \\
43   & cusp     & hooked        & LB       & attached ridge  &     2460      4100 &      933      1067 \\
44   & cusp     & -             & LB       & attached ridge  &     6151     20912 &     1059      1219 \\
45   & cusp     & -             & LB       & attached ridge  &     2460      4920 &     1268       945 \\
46   & cusp     & pointy        & LB       & attached ridge  &    16401      4100 &     1540      1256 \\
47   & cusp     & 3-tip         & LB       & attached ridge  &     9021     29933 &     1992      1310 \\
48   & cusp     & 2-tip         & LB       & attached ridge  &     5330     18042 &     2226      1364 \\
49   & sil.     & circular      & sharp    & near ridge      &     1599           &     2128      1046 \\
50   & cusp     & -             & LB       & near ridge      &     4510           &     3162       424 \\
51   & cusp     & long          & LB       & near ridge      &     3690      15171&     3322       393 \\
52   & cusp     & -             & LB       & near ridge      &      820      1230 &     3413       387 \\
53   & cusp     & -             & LB       & near ridge      &     2460           &     3334       881 \\
54   & cusp     & -             & LB       & near ridge      &     4100      5741 &     3117      1559 \\
\multicolumn{7}{l}{Pismis~24} \\                                                                   
1    & cusp     & rounded       & LB       & near ridge      &     1900           &      743      3395 \\
2    & cusp     & -             & LB       & near ridge      &     1500      4000 &      902      3349 \\
3    & cusp     & rounded/tail  & LB       & near ridge      &    12501     15001 &      915      3601 \\
4    & cusp     & long          & LB       & near ridge      &     2000      7751 &     1073      3658 \\
5    & cusp     & -             & LB       & near ridge      &      750      2250 &     1169      3723 \\
6    & cusp     & -             & LB       & near ridge      &     1900           &     1033      3777 \\
7    & cusp     & -             & LB       & near ridge      &     2125      4000 &     1463      3930 \\
8    & cusp     & -             & LB       & near ridge      &      499           &     1703      3786 \\
9    & cusp     & -             & LB       & near ridge      &     1775      3000 &      928      3231 \\
10   & cusp     & -             & LB       & attached        &     1800      3250 &      680      2981 \\
11   &fragment  & complex       & LB       & attached ridge  &    12001     32503 &      677      2908 \\
12   & cusp     & -             & LB       & attached        &     1850      5000 &      576      2818 \\
13   & sil.     & nodule        & sharp    & -               &      475      1900 &     1396      3076 \\
14   & cusp     & long/knotty   & LB       & attached        &     1450      6001 &      521      2432 \\
15   & cusp     & pointy        & LB       & attached        &     2500           &      624      2379 \\
16   &fragment  & complex       & LB       & attached ridge  &    15001     67506.&      408      2359 \\
17   & cusp     & -             & LB       & attached        &     1375           &      317      2558 \\
18   & sil.     & circular/halo & sharp    & near ridge      &     3250           &     2120      3049 \\
19   & cusp     & pointy        & LB       & attached ridge  &     2500      5500 &     2317      2843 \\
20   & sil.     & cusp-like     & fuzzy    & attached ridge  &     2500     25002 &     2148      2594 \\
  \multicolumn{7}{l}{IC~1590}\\                                                                                
1    &fragment  & complex       & LB       & isolated ?      &    55865    138192 &     1913      3220 \\
2    & cusp     & rounded       & LB       & attached frag. 1&     7645     17936 &     1853      2872 \\
3    &fragment  & rounded       & LB       & near frag. 1    &     8530           &     1805      3656 \\
4    &fragment  & rounded       & LB       & near frag. 1    &     6760     19700 &     1821      3800 \\
  \multicolumn{7}{l}{IC~2944\tablenotemark{c}} \\                                                                               
R1   &fragment  & complex       & sharp    & isolated        &    30603     48604 &     1465       837 \\
R2   &fragment  & complex       & sharp    & isolated        &    10441     18002 &      319       161 \\
R3   &fragment  & complex       & sharp    & isolated        &     9361     27002 &      976       201 \\
R11  &fragment  & complex       & sharp    & isolated        &     1980     15841 &      553       418 \\
R11B & cusp     & hooked        & LB       & attached frag.11&      774      1764 &      544       325 \\
R14  & sil.     & complex       & sharp    & -               &     4320      6373 &      857       845 \\
R15  & sil.     & complex       & sharp    & -               &     1800           &      963       831 \\
R16  & sil.     & circular      & sharp    & -               &     1980           &      913      1384 \\
1    & sil.     & rounded       & sharp    & attached frag. 1&     1980      2880 &      945      1228 \\
R17  & sil.     & circular      & sharp    & -               &     1368           &     1024      1340 \\
R18  &fragment  & complex       & fuzzy    & attached frag. 1&     3240      4500 &     1103      1437 \\
R19  & sil.     & circular      & sharp    & -               &     1260           &     1190      1490 \\
R20  & sil.     & rounded       & fuzzy    & -               &     2160           &     1370      1371 \\
R21  & sil.     & circular      & sharp    & -               &      990           &     1499      1219 \\
R22  & sil.     & complex       & sharp    & -               &     4500      5761 &     1438      1127 \\
2    & sil.     & nodule        & sharp    & -               &      720           &     1384      1171 \\
R23  & sil.     & long          & sharp    & -               &      900      3960 &     1434       891 \\
R24  & sil.     & long          & sharp    & -               &     1656      3060 &     1465       837 \\
R25  & sil.     & nodule        & sharp    & -               &     1008           &     1244       941 \\
R26  & sil.     & circular      & sharp    & -               &     1026           &     1183       793 \\
R27  & sil.     & circular      & sharp    & -               &     1044           &     1156       797 \\
R30/1& sil.     & irregular     & sharp    & -               &     5941      3240 &     1293       591 \\
R32  & sil.     & triangular    & sharp    & -               &     2160           &     1257       537 \\
R33  & sil.     & circular      & sharp    & -               &     1782           &      954       444 \\
R35  & sil.     & rounded       & sharp    & -               &      846           &     1061       686 \\
R36  & sil.     & complex       & sharp    & -               &     2520      3780 &      602       774 \\
3    & cusp     & -             & LB       & -               &     9541     14401 &      319       161 \\
R37  & sil.     & long          & sharp    & -               &     1458      3240 &      462       694 \\
4    & sil.     & nodule        & sharp    & -               &      900           &      483       608 \\
R38A & sil.     & circular      & sharp    & -               &     1350           &      481       635 \\
R38B & sil.     & knotty        & -        & -               &      630           &      436       636 \\
R39  & sil.     & nodule        & sharp    & -               &      882           &      742       160 \\
R40A & sil.     & nodule        & sharp    & -               &      450           &      451       267 \\
R40B & sil.     & nodule        & sharp    & -               &      468           &      457       252 \\
R40C & sil.     & nodule        & sharp    & -               &      522           &      468       235 \\
R41  & sil.     & nodule        & sharp    & -               &      558           &      119       186 \\
R42  & sil.     & nodule        & sharp    & -               &      540           &       48       206 \\
  \multicolumn{7}{l}{M~16} \\                                                                              
1    & sil.     & circular      & sharp    & near frag. 4    &      749           &     1383       736 \\
2    & sil.     & circular      & sharp    & near frag. 4    &     1498           &     1398       931 \\
3    & cusp     & oval          & LB       & near frag. 4    &      963      1904 &     1005      1509 \\
4    &fragment  & complex       & mix      & near ridge      &    25682           &      916      1307 \\
5    &cusp      & 3-point       & sharp    & attached frag. 4&     3210      6635 &      949      1478 \\
6    & cusp     & rounded       & sharp    & attached frag. 4&     1819      2568 &      882      1203 \\
7    & cusp     & rounded       & LB       & attached frag. 4&     1712      8132 &     1013      1152 \\
  \multicolumn{7}{l}{M~17} \\                                                                                  
1    & sil.     & -             & sharp    & attached frag. ?&      494       689 &     1572      4182 \\
2    &fragment  & -             & LB       & near ridge      &     1950      3900 &      705      3720 \\
3    & cusp     & -             & LB       & isolated        &      715           &     2676      3910 \\
4    &fragment  & -             & LB       & attached ridge ?&    28603     29903 &     3353      3736 \\
5    &fragment  & -             & LB       & attached ridge ?&    10921     23402 &     3362      3019 \\
6    & cusp     & long w/ tail  & LB       & near frag. 5    &     1170     14301 &     3866      2817 \\
7    & sil.     & circular      & mix      & isolated        &     1430      1690 &      972      2277 \\
8    & sil.     & hook-shaped   & sharp    & near frag. 6    &     1170           &     3930      3177 \\
9    & cusp     & -             & LB       & near cusp 10    &      325           &     3866      3564 \\
10   & cusp     & -             & LB       & near cusp 9     &      650           &     3866      3564 \\
11   & cusp     & rounded       & LB       & near cusps 9,10 &      845           &     4204      3900 \\
12   & sil.     & edge-on       & sharp    & near ridge      &     1170       195 &     2022      3844 \\
13   &fragment  & rounded       & LB       & isolated        &    16251     11246 &     1377      1550 \\
\enddata
\tablenotetext{a}{Linear sizes are determined using the distance estimates from Table~\ref{tab:clusters}.}
\tablenotetext{b}{These are the pixel positions in the following images: j8rh02011\_drz.fits (NGC~6530),
j8rh04011\_drz.fits (NGC~2467), j8rh05011\_drz.fits (Pismis~24), j8rh06011\_drz.fits (IC~1590), u57v0101r (IC~2944), 
u6dn3105m (M~16) and j8cw078b1i\_drz.fits (M~17)}
\tablenotetext{c}{The numbering scheme for these features follows \citet{Rei+03} except for a few fragments
(whose numbers are not preceded by the letter R) that were not identified by them.}
\tablecomments{Class: ``sil." stands for silhouette; Shape: when a `-' is seen in lieu of
the shape of a cusp it is because the cusp is tear-shaped.
Edges: ``LB" stands for limb-brightened. If
the edge is {\it not} LB then it can be sharp, fuzzy or a mix of both.}
\end{deluxetable}

\begin{deluxetable}{lllll}
\tabletypesize{\scriptsize}
\tablewidth{0pt}
\tablecaption{Sizes of proplyd and proplyd-like features.\label{tab:prop-summary}}
\tablehead{
\colhead{Region} & \colhead{Distance} & \colhead{Dark Sil.} & \colhead{Bright Cusps} & \colhead{Comments} \\
\colhead{}       & \colhead{(kpc)}    & \colhead{Size (AU)}  & \colhead{Chord (AU)}   & 
}
\startdata
&&&&\\
\multicolumn{5}{l}{From the literature.}\\
Orion Nebula\tablenotemark{a}   & 0.43      & 65-800     & 50-1700      & {\it Bona fide} proplyds mostly with central stars\\
NGC~3372\tablenotemark{b}& 2.3      & 3060       & 3750-6900    & Several found. No central stars.\\
M~20\tablenotemark{c}    & 1.68     & 470?       & 1070         & One found with disk, cusp and central star.\\
M~16\tablenotemark{d}& 2        & -          & 300-700      & Some with central stars.\\ 
NGC~3603\tablenotemark{e}& 6        & -          & 6000         & Several found.\\
M~8\tablenotemark{f}     & 1.8      & -          & 615          & One found with central star.\\
IC~2944                  & 1.8      & -          & 720-6400     & Many silhouettes, some circular.\\
&&&&\\
\multicolumn{5}{l}{This work.}\\
NGC~2467 & 4.1      &-           & 820-4500     & Few found no central stars.\\
Pismis~24& 2.5      &2400        & 500-2100     & One dark feature with halo, no star. \\
         &          &            &              & Few bright found no central stars.\\
M~16 & 2.14     &750-1500    & See above.   & Two found no central stars.\\
M~17     & 1.3      & 1170       & 325-845      & Few cusps found no central stars. \\
         &          &            &              & One disk with star; Fig.\ref{fig:m17_proplyd}.\\
\enddata
\tablenotetext{a}{\citet{Bal+00}; the largest cusp chords, $\sim$1700~AU, 
                  are those of proplyd 244-440 \citep{HO99} and 181-826 
\citep{Bal+05}.}

\tablenotetext{b}{\citet{Smi+03}}
\tablenotetext{c}{\citet{Yus+05}}
\tablenotetext{d}{\citet{Hes+96}; the associated stars are not in the middle of the cusps.}
\tablenotetext{e}{\citet{Bra+00}}
\tablenotetext{f}{\citet{Ste+98}}
\end{deluxetable}


\end{document}